\documentclass[pra,twocolumn,superscriptaddress,showpacs]{revtex4}

\usepackage[dvips]{graphics}
\usepackage{amssymb}
\usepackage{amsmath}
\usepackage{epsfig}
\usepackage{latexsym}
\usepackage{color}
\usepackage{rotating}
\usepackage{float}

\begin{document}

\newcommand{\boxedeqn}[1]{%
  \[\fbox{%
      \addtolength{\linewidth}{-2\fboxsep}%
     \addtolength{\linewidth}{-2\fboxrule}%
      \begin{minipage}{\linewidth}%
     \begin{equation}#1\end{equation} %
     \end{minipage}%
   }\]%
}

\title{Two-mode squeezed states in cavity optomechanics \\ via engineering of a single reservoir}
\author{M. J. Woolley} 
\affiliation{School of Engineering and Information Technology, UNSW Canberra, ACT, 2600, Australia}
\author{A. A. Clerk }
\affiliation{Department of Physics, McGill University, Montr\'{e}al, QC, H3A 2T8, Canada}

\begin{abstract}
We study theoretically a three-mode optomechanical system where two mechanical oscillators are independently coupled to a single cavity mode. By optimized two-tone or four-tone driving of the cavity one can prepare the mechanical oscillators in an entangled two-mode squeezed state, even if they start in a thermal state. The highly-pure, symmetric steady-state achieved allows the optimal fidelity of standard continuous-variable teleportation protocols to be achieved. In contrast to other reservoir engineering approaches to generating mechanical entanglement, only a single reservoir is required to prepare the highly-pure entangled steady-state, greatly simplifying experimental implementation. The entanglement may be verified via a bound on the Duan inequality obtained from the cavity output spectrum. A similar technique may be used for the preparation of a highly-pure two-mode squeezed state of two cavity modes, coupled to a common mechanical oscillator.    
\end{abstract}


\maketitle

\section{Introduction}
\label{sec:intro}
The generation and detection of entangled states of macroscopic mechanical oscillators is an outstanding task in the study of mechanical systems in the quantum regime \cite{woolley:review}. There exist a number of proposals for the generation of such states \cite{mancini,schmidt,tan,woolley:twomode,bowen}. Perhaps most promising amongst these are approaches based on reservoir engineering \cite{muschik,lukin,wang:coolBogoliubov,meystre}, whereby the dissipation is engineered such that the steady-state of the dissipative dynamics is the desired target state. These proposals are highly attractive from an experimental point-of-view, requiring relatively minor modifications of existing experimental configurations \cite{schwab1,teufel,massel,schwab4}. 

Here we propose an approach for generating highly-pure, highly-entangled two-mode squeezed states of two mechanical oscillators via coupling to a driven cavity mode. The two-mode squeezed state is the simultaneous vacuum of two non-local bosonic operators (so-called Bogoliubov modes) \cite{bruus}. Hence it is possible to prepare a mechanical two-mode squeezed state by cooling these two modes. This can be achieved using two independent reservoirs (see, e.g. Refs.~\onlinecite{muschik,wang:coolBogoliubov,meystre}). Here however, we show that the same goal can be achieved using just a \emph{single} reservoir. By making the Bogoliubov modes non-degenerate they will couple to different frequency components of a single reservoir (that is, the damped cavity). Since the Bogoliubov transformation preserves the difference in number operators, this simply corresponds to a frequency difference of the two mechanical oscillators. Viewed differently, one can say that we are exploiting the coherent dynamics of the mechanical oscillators to effectively cool both Bogoliubov modes. Ultimately, our protocol simply involves appropriately weighted and detuned two-tone or four-tone driving of the coupled cavity mode. Note that the entanglement of oscillators interacting with a common reservoir has also been discussed in Ref.~\onlinecite{ankerhold}.

The approach we take here may be regarded as the coherent feedback \cite{kerckhoff} analogue of our measurement-based approach to the same task \cite{woolley:twomode}, and is related to a recent proposal for the preparation of a quantum squeezed state of a single mechanical oscillator \cite{kronwald}. As compared with the measurement-based approach for entanglement generation in Ref.~\onlinecite{woolley:twomode}, the purity of the steady-state achieved here is greater and the implementation is greatly simplified. Further, in contrast to the proposal of Ref.~\onlinecite{wang:coolBogoliubov}, the steady-state is a (highly-pure) two-mode squeezed state, rather than an entangled mixed state. Therefore, using the steady-state as an EPR channel, the optimal teleportation fidelity for a given amount of entanglement can be achieved via the standard continuous-variable teleportation protocol, without the need for additional local operations \cite{braunstein,adessoilluminati}. In contrast to the proposal of Ref.~\onlinecite{meystre} (also described in the supplement to Ref.~\onlinecite{wang:coolBogoliubov}) the two-mode squeezed state is generated here using only one, rather than two, auxiliary cavity modes, simplifying the experimental implementation. 

Reservoir engineering has earlier been the subject of significant theoretical study in the context of optical and atomic systems \cite{cirac93,zoller,plenio:dissipativeentanglementlight,parkins:dissipativeentanglementatoms,muschik,dalla,he}, culminating in the experimental demonstration by Krauter and co-workers of the entanglement of atomic ensembles \cite{krauter:experimentaldissipativeentanglement}. The utility of reservoir engineering has also been shown with two-level systems, with demonstrations of superconducting qubit state control \cite{murch}, as well as entanglement in both trapped ion \cite{wineland} and superconducting \cite{shankar} systems. In other work pertaining to mechanical entanglement, the entanglement of mechanical motion with a microwave field has been demonstrated \cite{lehnert}. Earlier, entanglement of phonons at the single-quantum level was demonstrated \cite{diamonds}, as well as the entanglement of motional degrees of freedom of trapped ions \cite{ions}. 

Here, in Sec.~\ref{sec:system} we introduce the multimode optomechanical system that we shall study. Sec.~\ref{sec:reservoirengineer} describes approaches that one may take to reservoir engineering in this system, while in Sec.~\ref{sec:implementation} we describe how these strategies could be implemented in our system. In Sec.~\ref{sec:adiabatic} we consider the adiabatic limit, in which the cavity responds rapidly to the mechanical motion, and obtain analytical expressions for the entanglement, purity and teleportation fidelity possible with the steady-state. Sec.~\ref{sec:full} gives an analysis of the full linearized system, including the effects of counter-rotating terms and the possibility of instability. In Sec.~\ref{sec:expt} we derive a bound on the entanglement based on the cavity output spectrum. Sec.~\ref{sec:cavity} provides an analysis of the three-mode optomechanical system composed of two cavity modes coupled to a single mechanical oscillator, and demonstrates that the same physics can be realized in this system. 

\section{System and Hamiltonian}
\label{sec:system}

\setlength{\abovecaptionskip}{-10pt }

\begin{figure}[ht]
\begin{center}
\includegraphics[scale=0.38]{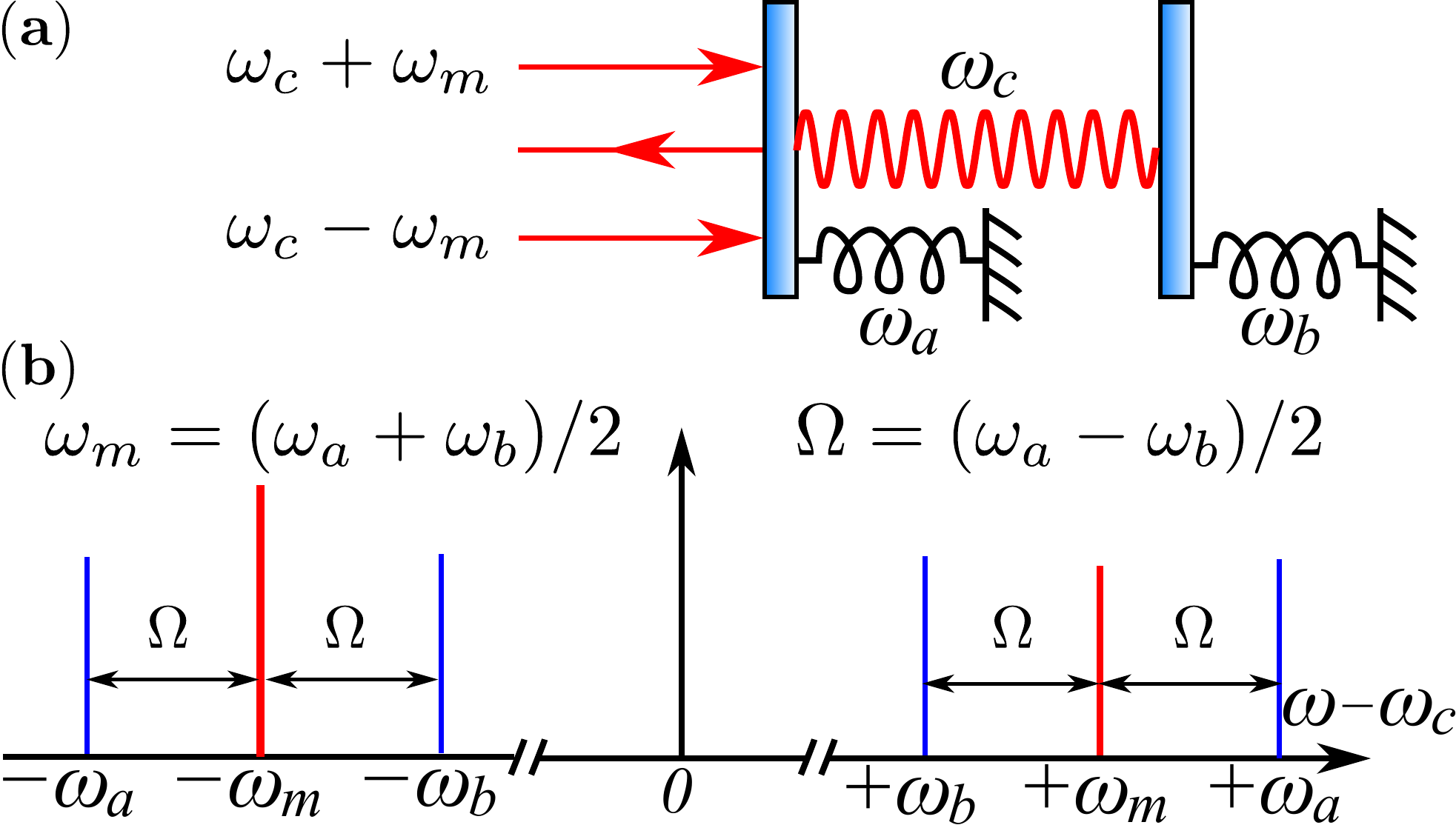}
\end{center}
\caption{  (Color online) (a) The system studied consists of two mechanical oscillators, each coupled to a common cavity (or circuit) mode. (b) Frequencies in this system defined with respect to the cavity resonance frequency $\omega_c$. The blue lines indicate the standard mechanical sidebands, at $\pm \omega_a$ and $\pm \omega_b$. If the single-photon optomechanical coupling rates ($g_a$ and $g_b$) are equal, the required Hamiltonian (\ref{eq:HamGen1}) can be realised using only two cavity drive detunings ($\pm \omega_m$), indicated by the vertical red lines. } 
\label{fig:optodualmechanics}
\end{figure}

The system, see Fig.~\ref{fig:optodualmechanics}(a), is composed of two mechanical oscillators, with resonance frequencies $\omega_a$ and $\omega_b$, each independently, dispersively coupled (with strengths $g_a$ and $g_b$, respectively) to a common cavity mode having resonance frequency $\omega_c$. The Hamiltonian is
\begin{eqnarray}
\mathcal{\hat{H}} & = & \omega_a \hat{a}^\dagger \hat{a} + \omega_b \hat{b}^\dagger \hat{b} + \omega_c \hat{c}^\dagger \hat{c} + g_a \left( \hat{a}+\hat{a}^\dagger \right) \hat{c}^\dagger \hat{c} \nonumber \\
& &  + g_b ( \hat{b}+\hat{b}^\dagger ) \hat{c}^\dagger \hat{c} + \hat{H}_{\rm drive} + \hat{H}_{\rm diss} , \label{eq:HamOne}
\end{eqnarray}
where $\hat{a}$ and $\hat{b}$ denote mechanical mode lowering operators, $\hat{c}$ denotes the electromagnetic mode lowering operator, and $\hat{H}_{\rm drive}$ accounts for driving of the electromagnetic mode. The term $\hat{H}_{\rm diss}$ accounts for dissipation, with the modes subject to damping at rates $\gamma_a$, $\gamma_b$ and $\kappa$, respectively. The system dynamics, within the usual approximations \cite{walls}, are described by the master equation:
\begin{eqnarray}
\dot{\rho} & = & -i [ \hat{\mathcal{H}}', \rho ] + \gamma_a (\bar{n}_a + 1 ) \mathcal{D}[ \hat{a} ] \rho + \gamma_a \bar{n}_a \mathcal{D} [ \hat{a}^\dagger ] \rho \nonumber \\
& & + \gamma_b (\bar{n}_b + 1 ) \mathcal{D}[ \hat{b} ] \rho + \gamma_b \bar{n}_b \mathcal{D} [ \hat{b}^\dagger ] \rho + \kappa \mathcal{D} [ \hat{c} ] \rho , \label{eq:MEOriginal}
\end{eqnarray}
with the Hamiltonian $\hat{\mathcal{H}}' = \hat{\mathcal{H}} - \hat{H}_{\rm diss}$ and the dissipative superoperator $ \mathcal{D} [ \hat{s} ] \rho = \hat{s} \rho \hat{s}^\dagger - \frac{1}{2} \hat{s}^\dagger \hat{s} \rho - \frac{1}{2} \rho \hat{s}^\dagger \hat{s} $. 

\section{Reservoir Engineering Strategies}
\label{sec:reservoirengineer}
Given the two mechanical oscillator modes $\hat{a}$ and $\hat{b}$, one can introduce mechanical two-mode Bogoliubov operators in some rotating frame,
\begin{subequations}
\begin{eqnarray}
\hat{\beta}_1 & = & \hat{a} \cosh r + \hat{b}^\dagger \sinh r , \label{eq:bog1} \\
\hat{\beta}_2 & = & \hat{b} \cosh r + \hat{a}^\dagger \sinh r . \label{eq:bog2}
\end{eqnarray}
\end{subequations}
Typically, the choice of rotating frame is set by the resonance frequencies of the system, though here the rotating frame shall be defined with respect to 
\begin{equation}
\hat{H}_0 = (\omega_a - \Omega ) \hat{a}^\dagger \hat{a} + (\omega_b + \Omega ) \hat{b}^\dagger \hat{b} + \omega_c \hat{c}^\dagger \hat{c} , \label{eq:framedef}
\end{equation}
with the choice of $\Omega$ to be specified in Sec.~\ref{sec:implementation}. Note that in this frame the collective mechanical quadratures, to be defined precisely in Eqs.~(\ref{eq:collectivequad1}) and (\ref{eq:collectivequad2}), are rotating in a non-trivial manner. In particular,
\begin{subequations}
\begin{eqnarray}
\hat{X}_+ & = & \frac{1}{2} \left[ \hat{a} e^{+i(\omega_a - \Omega )t} + \hat{a}^\dagger e^{-i(\omega_a - \Omega )t} \right. \nonumber \\
& & \left. + \hat{b} e^{+i(\omega_b + \Omega )t} + \hat{b}^\dagger e^{-i(\omega_b + \Omega )t} \right] , \\
\hat{P}_- & = & -\frac{i}{2} \left[ \hat{a} e^{+i(\omega_a - \Omega )t} - \hat{a}^\dagger e^{-i(\omega_a - \Omega ) t} \right. \nonumber \\
& & \left. + \hat{b} e^{+i(\omega_b + \Omega )t} - \hat{b}^\dagger e^{-i(\omega_b + \Omega )t} \right] .
\end{eqnarray}
\end{subequations}

In any case, the two-mode squeezed state is defined as $\left| r \right\rangle_2 = \hat{S}_2(r) \left| 0,0 \right\rangle$ where 
\begin{equation}
\hat{S}_2(r) \equiv \exp [ r ( \hat{a} \hat{b} - \hat{a}^\dagger \hat{b}^\dagger ) ] , \label{eq:TMSoperator}
\end{equation}
is the two-mode squeezing operator with squeezing parameter $r$ \cite{gerry}. Starting from $\hat{a} \left| 0,0 \right\rangle = 0$ and $\hat{b} \left| 0,0 \right\rangle = 0$ it is straightforward to show that $[ \hat{S}_2(r) \hat{a} \hat{S}^\dagger_2(r) ] \left| r \right\rangle_2 = 0$ and that $[ \hat{S}_2(r) \hat{b} \hat{S}^\dagger_2(r) ] \left| r \right\rangle_2 = 0$. However, $\hat{\beta}_1 = \hat{S}_2(r) \hat{a} \hat{S}^\dagger_2(r)$ and $\hat{\beta}_2 = \hat{S}_2(r) \hat{b} \hat{S}^\dagger_2(r)$, and therefore, the ground state of $\hat{\beta}_1$ and $\hat{\beta}_2$ is the two-mode squeezed state with squeezing parameter $r$. Our goal then is to engineer the driving Hamiltonian of (\ref{eq:HamOne}) such that the steady-state of (\ref{eq:MEOriginal}) results in the $\hat{\beta}_i$ modes being cooled to their ground state, implying two-mode squeezing of the mechanical oscillators.  

One method to achieve this is to use two cavity modes to independently cool the Bogoliubov modes \cite{muschik,wang:coolBogoliubov,meystre}. Both Bogoliubov modes are independently coupled to a cavity mode with a beam-splitter-like interaction, i.e. $\hat{\beta}^\dagger_i \hat{c}_i + {\rm h.c.}$ ($i=1,2$), an interaction which can be used to cool the Bogoliubov modes \cite{woolley:nanomechanicalsqueezing}. While such an approach can be effective, from a practical point of view it would be highly advantageous if this could be achieved using only a single cavity mode.

A seemingly simple way of using only a single reservoir would be to couple the cavity to one of the Bogoliubov modes, say $\hat{\beta}_1$, and then couple $\hat{\beta}_1$ to the other Bogoliubov mode ($\hat{\beta}_2$) via an all-mechanical contribution to the Hamiltonian of the form $\hat{\beta}^\dagger_1 \hat{\beta}_2 + {\rm h.c.}$, again a beam-splitter-like interaction. This interaction will allow $\hat{\beta}_2$ to be cooled (by swapping quanta into $\hat{\beta}_1$) even though it is not directly coupled to the cooling reservoir. While conceptually simple, the requisite mechanical interaction would be difficult to realise, requiring the direct coupling of the mechanical oscillators. Such a Hamiltonian was considered in Ref.~\onlinecite{yamamoto}, albeit without a physical implementation specified. 

A third approach, which we will pursue here, is to couple the cavity to the sum of Bogoliubov modes, $\hat{\beta}_{\rm sum} \equiv (\hat{\beta}_1 + \hat{\beta}_2)/\sqrt{2}$, and then couple the sum mode to the difference mode via an all-mechanical Hamiltonian contribution of the form $\hat{\beta}^\dagger_{\rm sum} \hat{\beta}_{\rm diff} + {\rm h.c.}$, where $\hat{\beta}_{\rm diff} \equiv (\hat{\beta}_1 - \hat{\beta}_2)/\sqrt{2}$. Again, the swap interaction allows $\hat{\beta}_{\rm diff}$ to be cooled even though it is not directly coupled to the cavity. Cooling both $\hat{\beta}_{\rm sum}$ and $\hat{\beta}_{\rm diff}$ is equivalent to cooling both $\hat{\beta}_1$ and $\hat{\beta}_2$, since $\langle \hat{\beta}^\dagger_{\rm sum} \hat{\beta}_{\rm sum} \rangle + \langle \hat{\beta}^\dagger_{\rm diff} \hat{\beta}_{\rm diff}  \rangle = \langle  \hat{\beta}^\dagger_{1} \hat{\beta}_{1}  \rangle + \langle \hat{\beta}^\dagger_{2} \hat{\beta}_{2} \rangle $. While this approach again seems to involve the realization of a challenging interaction between Bogoliubov modes, this is not the case. The beam-splitter interaction here takes the simple form: $\hat{\beta}^\dagger_1 \hat{\beta}_1 - \hat{\beta}^\dagger_2 \hat{\beta}_2 = \hat{a}^\dagger \hat{a} - \hat{b}^\dagger \hat{b}$. Thus, one does not require a direct interaction between the mechanical oscillators, but rather just a difference in their resonance frequencies. This is the key insight that allows one to appropriately engineer the reservoir via a single cavity mode. 

As noted in Sec.~\ref{sec:intro}, we may take another perspective on this third approach. By introducing a frequency difference between the two mechanical oscillators, we are breaking the degeneracy of the Bogoliubov modes $1$ and $2$, since the Bogoliubov transformation preserves the number operator difference. Consequently, the Bogoliubov modes couple to different frequency components of the reservoir. Due to the finite bandwidth of the cavity it effectively functions as two independent reservoirs such that both Bogoliubov modes are cooled. 

We thus have that the desired Hamiltonian, in terms of the Bogoliubov modes defined in Eqs.~(\ref{eq:bog1}) and (\ref{eq:bog2}), is
\begin{eqnarray}
\mathcal{\hat{H}} & = & \Omega \left( \hat{\beta}^\dagger_1 \hat{\beta}_1 - \hat{\beta}^\dagger_2 \hat{\beta}_2 \right) + \mathcal{G} \left[ \left( \hat{\beta}^\dagger_1 + \hat{\beta}^\dagger_2 \right) \hat{c} + {\rm h.c.} \right] \nonumber \\
& & + \hat{H}_{\rm diss} , \label{eq:HamCoolAsBogoliubov} 
\end{eqnarray}
where $\Omega$ is an effective oscillation frequency and $\mathcal{G}$ is an effective optomechanical coupling. In terms of the original mechanical annihilation operators, Eq.~(\ref{eq:HamCoolAsBogoliubov}) is
\begin{eqnarray}
\hat{\mathcal{H}} & = & \Omega \left( \hat{a}^\dagger \hat{a} - \hat{b}^\dagger \hat{b} \right) + G_+ \left[ ( \hat{a} + \hat{b} ) \hat{c} + {\rm h.c.} \right] \nonumber \\
& & + G_- \left[ ( \hat{a} + \hat{b} ) \hat{c}^\dagger + {\rm h.c.} \right] + \hat{H}_{\rm diss} . \label{eq:HamGen1} 
\end{eqnarray}
The optomechanical couplings in Eqs.~(\ref{eq:HamCoolAsBogoliubov}) and (\ref{eq:HamGen1}) are related by 
\begin{subequations}
\begin{eqnarray}
\mathcal{G} & \equiv & \sqrt{ G^2_- - G^2_+ } , \label{eq:scriptGdef} \\ 
\tanh r & \equiv & G_+/G_-, \label{eq:r} 
\end{eqnarray}
\end{subequations}
with $r$ being the squeezing parameter entering in the definitions of the Bogoliubov modes in Eqs.~(\ref{eq:bog1}) and (\ref{eq:bog2}). 

Note that if $G_+ = G_-$ in Eq.~(\ref{eq:HamGen1}) then we recover the Hamiltonian required for a two-mode back-action-evading measurement of the mechanical oscillators \cite{woolley:twomode}, in which two collective mechanical quadratures commute with the system Hamiltonian. For $G_+ \neq G_-$, the back-action-evasion is lost, but now there is a back-action that may be regarded as a coherent feedback process. This enables two-mode squeezing without an explicit measurement. 
 
\section{Implementation} 
\label{sec:implementation}
The Hamiltonian (\ref{eq:HamGen1}) is readily implemented in conventional cavity optomechanics setups. We shall focus on the regime $| G_+ | < | G_- |$ such that the dynamics corresponding to (\ref{eq:HamGen1}) are stable. If the single-photon optomechanical coupling rates in (\ref{eq:HamOne}) are equal ($g_a = g_b$) then just two cavity drives are required to realize (\ref{eq:HamGen1}). If they are unequal ($g_a \neq g_b$) then four cavity drives are required. Of course, if $g_a \sim g_ b$, we can approximately realise (\ref{eq:HamGen1}) with only two cavity drives and still generate useful entanglement in the steady-state. We consider each of these cases in turn. 

\subsection{Two-tone driving}
If the single-photon optomechanical coupling rates are equal, we require cavity driving tones at $\omega_c \pm \omega_m$, where $\omega_m = (\omega_a + \omega_b)/2$ is the average of the two mechanical frequencies, i.e.: 
\begin{equation}
\hat{H}_{\rm drive} = \left( \mathcal{E}^*_+ e^{+i \omega_m t} + \mathcal{E}^*_- e^{-i \omega_m t} \right) e^{+i \omega_c t} \hat{c} + \textrm{h.c.} \label{eq:twotone}
\end{equation}
This situation is depicted in Fig.~\ref{fig:optodualmechanics}(b). The driving tones must be applied with a fixed relative phase. Working in an interaction picture defined with respect to the $\hat{H}_0 = \omega_m (\hat{a}^\dagger \hat{a} + \hat{b}^\dagger \hat{b}) + \omega_c \hat{c}^\dagger \hat{c}$, one finds the effective Hamiltonian to be given by Eq.~(\ref{eq:HamGen1}) where 
\begin{equation}
\Omega = ( \omega_a - \omega_b )/2,
\end{equation}
and the many-photon optomechanical couplings are 
\begin{equation}
G_{\pm} = (g_a+g_b) \bar{c}_{\pm}/2, \label{eq:optomechTwo}
\end{equation}
with $\bar{c}_{\pm}$ denoting the (assumed real) steady-state amplitudes of the fields at the driven sidebands, 
\begin{equation}
\bar{c}_{\pm} \equiv \langle \hat{c}_{\pm} \rangle_{\rm ss} = \frac{ i \mathcal{E}_{\pm} }{ \pm i \omega_m - \kappa/2 } .
\end{equation}
The details of this derivation are given in App.~\ref{sec:derivations}. It relies on the assumptions that we are working in the resolved-sideband regime ($\omega_a, \omega_b \gg \kappa$) and that the driving strengths $\mathcal{E}_{\pm}$ are large. The former assumption allows us to discard time-dependent contributions to the Hamiltonian (\ref{eq:HamGen1}), while the latter assumption allows us to linearize the optomechanical interaction. Note that the ratio $G_+/G_-$ shall be referred to here as the \emph{drive asymmetry} since it is set by the ratio of the cavity drives on either side of the cavity resonance frequency. 

If the single-photon optomechanical couplings are unequal the two-tone cavity driving cannot yield the complete matching of oscillator $a$ and $b$ sideband processes required in the ideal Hamiltonian of Eq.~(\ref{eq:HamGen1}). Instead there will be additional contributions to the Hamiltonian (\ref{eq:HamGen1}), given by
\begin{eqnarray}
\hat{H}_{\rm m} & = & G^{\rm m}_+ \left[ ( \hat{a} - \hat{b} ) \hat{c} + {\rm h.c.}  \right] \nonumber \\
& & - G^{\rm m}_- \left[ ( \hat{a}^\dagger - \hat{b}^\dagger ) \hat{c} + {\rm h.c.} \right] , \label{eq:HamBetter}
\end{eqnarray}
where the \emph{effective coupling imperfections} are
\begin{equation}
G^{\rm m}_{\pm} = \pm (g_a-g_b) \bar{c}_{\pm}/2. \label{eq:coupleimperfect2}  
\end{equation}
In the two-tone driving case the imperfection is due to the mismatch in the single-photon optomechanical coupling rates. 

\subsection{Four-tone driving}
\label{sec:fourtonedriving}
\begin{figure}[ht]
\begin{center}
\includegraphics[scale=0.38]{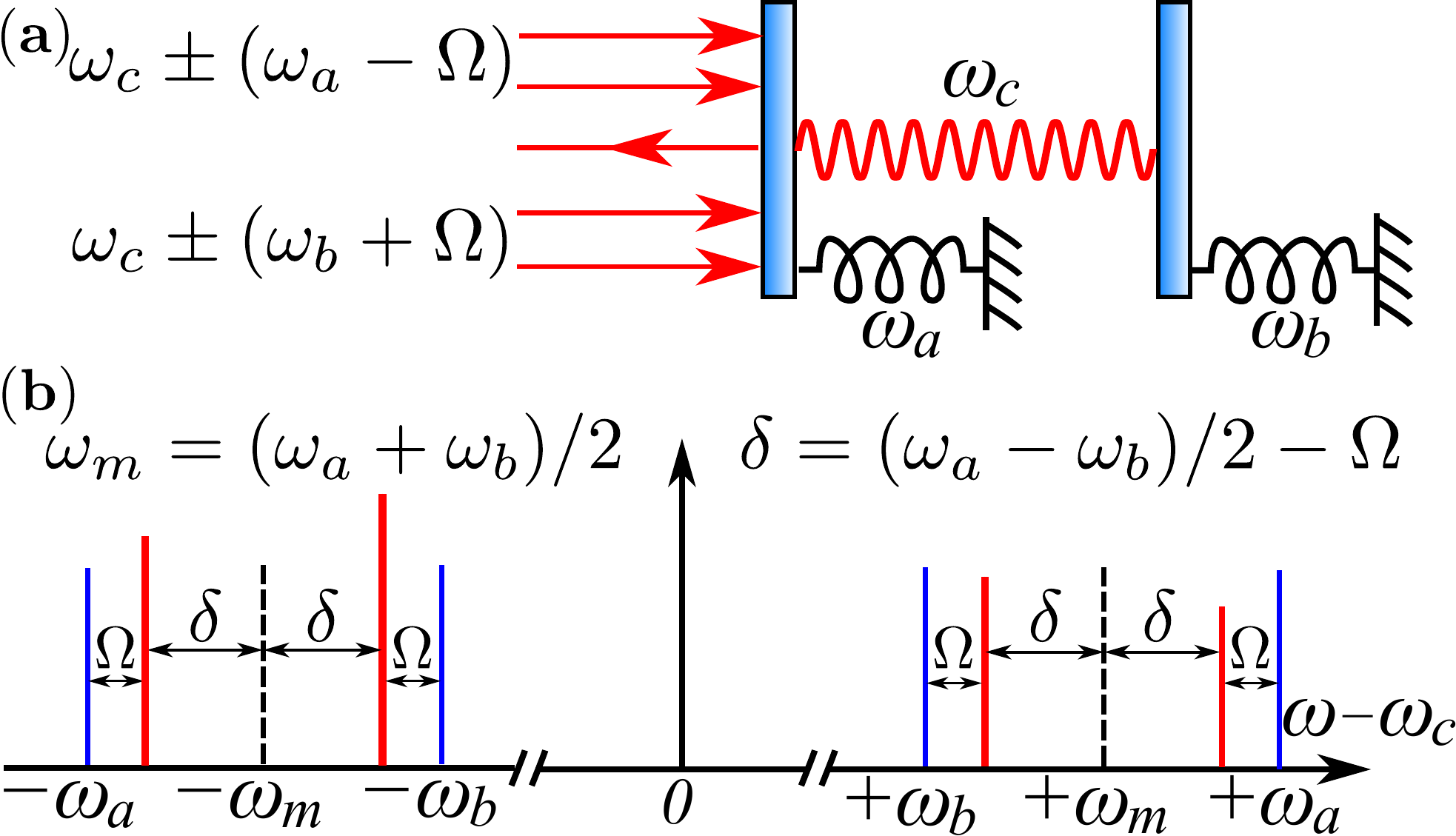}
\end{center}
\caption{  (Color online) (a) The three-mode optomechanical system, as in Fig.~\ref{fig:optodualmechanics}, under four-tone driving. (b) Frequencies in this system defined with respect to the cavity resonance frequency $\omega_c$. The blue lines indicate the standard mechanical sidebands, at $\pm \omega_a$ and $\pm \omega_b$. If the single-photon optomechanical coupling rates are unequal, the required Hamiltonian (\ref{eq:HamGen1}) can be realised using four cavity driving frequencies, $\pm ( \omega_a - \Omega), \pm (\omega_b + \Omega)$, indicated by vertical red lines. } 
\label{fig:optomechanics4tone}
\end{figure}

The Hamiltonian (\ref{eq:HamGen1}) involves four sideband processes; the up-conversion and down-conversion of drive photons via the absorption (or emission) of quanta from (or to) the mechanical oscillator $a$ or $b$. The realisation of (\ref{eq:HamGen1}) requires a balance of the rates at which these processes take place. By using four cavity driving tones, one tone associated with each sideband process, the balancing of the rates of these processes is possible even if the single-photon optomechanical couplings are unequal (see Fig.~\ref{fig:optomechanics4tone}). These driving tones are applied with a detuning of $\Omega$ from the mechanical sidebands, at $\omega_c \pm (\omega_a - \Omega )$ and $\omega_c \pm (\omega_b + \Omega )$, as depicted in Fig.~\ref{fig:optomechanics4tone}. The appropriate Hamiltonian contribution is
\begin{eqnarray}
\hat{H}_{\rm drive} & = & e^{+i\omega_c t} \hat{c} \left( \mathcal{E}^*_{1+} e^{+i (\omega_a - \Omega) t} + \mathcal{E}^*_{2+} e^{+i (\omega_b + \Omega) t} \right. \nonumber \\
& & \left. + \mathcal{E}^*_{1-} e^{-i (\omega_a - \Omega) t} + \mathcal{E}^*_{2-} e^{-i(\omega_b + \Omega) t} \right) + \rm{h.c.} \nonumber \\
& & \label{eq:fourdrives}
\end{eqnarray}
The steady-state amplitudes at the driven sidebands are denoted by $\bar{c}_{k\pm}$ ($k=1,2$), with
\begin{equation}
\bar{c}_{k\pm} \equiv \langle \hat{c}_{k \pm } \rangle_{\rm ss} = \frac{ i \mathcal{E}_{k\pm} }{ \pm i \omega_k - \kappa/2 } , \label{eq:SBamplitudes}
\end{equation}
where we have introduced the notation for the drive detunings
\begin{subequations}
\begin{eqnarray}
\omega_{1} & \equiv & \omega_a - \Omega , \label{eq:shorthand1} \\ 
\omega_2 & \equiv & \omega_b + \Omega . \label{eq:shorthand2}
\end{eqnarray}
\end{subequations}
Then we demand that the driving strengths are ``matched'', meaning that
\begin{equation}
\frac{ \bar{c}_{1\pm}}{ \bar{c}_{2\pm}} = \frac{g_b}{g_a} . \label{eq:matchingcondition}
\end{equation}
That is, we require that the two steady-state amplitudes (i.e. drives) on the same side of the cavity resonance frequency have an asymmetry set by the optomechanical coupling asymmetry. With the condition (\ref{eq:matchingcondition}) satisfied, and working in an interaction picture defined with respect to the Hamiltonian (\ref{eq:framedef}), the system can again be described by the Hamiltonian (\ref{eq:HamGen1}), now with the (assumed real) many-photon optomechanical coupling rates 
\begin{equation}
G_{\pm} = \left( g_a \bar{c}_{1\pm} + g_b \bar{c}_{2\pm} \right)/2 . \label{eq:GPlusMinus}
\end{equation}
Again, the details of the derivation are left to App.~\ref{sec:derivations}. Imprecision in the matching condition (\ref{eq:matchingcondition}) gives additional contributions to the Hamiltonian (\ref{eq:HamGen1}), of the form of Eq.~(\ref{eq:HamBetter}), but now with
\begin{eqnarray}
G^{\rm m}_{\pm} & = & \pm \left( g_a \bar{c}_{1\pm} - g_b \bar{c}_{2\pm} \right)/2 . \label{eq:Gmismatched}  
\end{eqnarray}
In this case the effective coupling imperfection arises from the drives not being weighted precisely according to the condition (\ref{eq:matchingcondition}).

The cavity drive frequencies should be set such that $\Omega$ satisfies the following conditions: 
\begin{subequations}
\begin{eqnarray}
\Omega & \gg & \gamma , \label{eq:OmegaGamma} \\
\Omega & \ll & (\omega_a - \omega_b)/2 - \gamma \label{eq:unwantedSB} . 
\end{eqnarray}
\end{subequations}
Condition (\ref{eq:OmegaGamma}) ensures that the sum and difference Bogoliubov modes are sufficiently coupled; c.f. Eq.~(\ref{eq:HamCoolAsBogoliubov}). The condition (\ref{eq:unwantedSB}) ensures that the unwanted sideband processes have a negligible effect on the system dynamics. It is some times convenient to refer to the drive frequencies via their detunings from the centre of the two mechanical sidebands, given by
\begin{equation}
\delta \equiv (\omega_a - \omega_b)/2 - \Omega . \label{eq:deltaDef}
\end{equation}

It is interesting to note that with the driving condition $\bar{c}_{1+} = \bar{c}_{1-}$, in addition to (\ref{eq:matchingcondition}), we could realize a two-mode back-action-evading measurement of the mechanical oscillators \emph{irrespective of the coupling asymmetry}, and so generalize the results of Ref.~\onlinecite{woolley:twomode}. 

\section{Adiabatic limit}
\label{sec:adiabatic}
We first consider the dynamics of the system governed by (\ref{eq:HamGen1}) in the adiabatic limit, where the cavity responds rapidly to the mechanical motion; that is, where $\kappa > \Omega , G_{\pm}$ (but still in the regime where $\omega_a, \omega_b \gg \kappa$). In this limit we eliminate the cavity mode, obtaining an effective description for the mechanical modes alone. This adiabatic limit will simplify the analysis and thus provide insight into our mechanism; it will also prove to be a useful regime for the task of mechanical entanglement generation. 

We stress that the Hamiltonian (\ref{eq:HamGen1}) applies both to the case of equal single-photon optomechanical couplings and two-tone driving, and to the case of unequal single-photon optomechanical couplings with matched four-tone driving (though $\Omega$ is determined differently in each case). Further, the imperfections (asymmetry in couplings in the former case and mismatch in driving conditions in the latter case) are both described by the Hamiltonian (\ref{eq:HamBetter}). 

In this adiabatic limit the cavity annihilation operator is given by $ \hat{c} = - 2i \mathcal{G} ( \hat{\beta}_1 + \hat{\beta}_2)/\kappa $. Substituting this into the dissipative terms of the master equation (\ref{eq:MEOriginal}), the adiabatically-eliminated master equation is
\begin{eqnarray}
\dot{\rho} & = & - i \Omega [ \hat{\beta}^\dagger_1 \hat{\beta}_1 - \hat{\beta}^\dagger_2 \hat{\beta}_2, \rho ] + \gamma_a \left( \bar{n}_a + 1 \right) \mathcal{D} [ \hat{a} ] \rho \nonumber \\
& & + \gamma_a \bar{n}_a \mathcal{D} [ \hat{a}^\dagger ] \rho + \gamma_b \left( \bar{n}_b + 1 \right) \mathcal{D} [ \hat{b} ] \rho + \gamma_b \bar{n}_b \mathcal{D} [ \hat{b}^\dagger ] \rho \nonumber \\
& & + \Gamma \mathcal{D} [ \hat{\beta}_1 + \hat{\beta}_2 ] \rho , \label{eq:MEShorter}
\end{eqnarray}
with the optomechanical damping rate,
\begin{equation}
\Gamma \equiv \gamma \mathcal{C} \equiv \frac{4\mathcal{G}^2}{\kappa } , \label{eq:optomechanicaldamping}
\end{equation}
where $\mathcal{G}$ is the effective optomechanical coupling introduced in Eq.~(\ref{eq:scriptGdef}), and $\mathcal{C}$ is the corresponding cooperativity parameter. Now, the steady-state of Eq.~(\ref{eq:MEShorter}) is easily obtained, and its entanglement and purity metrics readily calculated. 

In view of Eq.~(\ref{eq:MEShorter}), an alternative interpretation of the cooling of both Bogoliubov modes is possible. In the limit $\gamma/\Omega \rightarrow 0$, the terms containing $\hat{\beta}^\dagger_1 \hat{\beta}_2$ and $\hat{\beta}^\dagger_2 \hat{\beta}_1$ will rapidly average away, meaning that Eq.~(\ref{eq:MEShorter}) will be equivalent to having independent dissipation of modes $\hat{\beta}_1$ and $\hat{\beta}_2$. Physically this corresponds to the dissipation of $\hat{\beta}_1$ and $\hat{\beta}_2$ being due to distinct modes of the reservoir. 

\subsection{Entanglement}
The case of symmetric mechanical damping ($\gamma_a , \gamma_b = \gamma$) and symmetric thermal occupation of the mechanical baths ($\bar{n}_a , \bar{n}_b = \bar{n}$) allows simple analytical results for the steady-state second moments to be obtained. The assumption of equal thermal occupations is reasonable for most experimental situations, while it turns out that our results are not sensitive to unequal mechanical damping rates provided that they are both small. The simplest two-mode, continuous-variable entanglement criterion is provided by the Duan inequality \cite{duan}. It is expressed in terms of collective quadrature operators, defined by
\begin{subequations}
\begin{eqnarray}
\hat{X}_{\pm} & = & ( \hat{X}_a \pm \hat{X}_b )/\sqrt{2} , \label{eq:collectivequad1} \\
\hat{P}_{\pm} & = & ( \hat{P}_a \pm \hat{P}_b )/\sqrt{2} , \label{eq:collectivequad2}
\end{eqnarray}
\end{subequations}
where we have introduced the usual quadratures for each oscillator mode, 
\begin{equation}
\hat{X}_s = (\hat{s} + \hat{s}^\dagger )/\sqrt{2}, \ \ \hat{P}_s = -i(\hat{s} - \hat{s}^\dagger )/\sqrt{2}. \label{eq:usualquadratures}
\end{equation}
Then the Duan criterion tells us that a Gaussian state for which
\begin{equation}
\langle \hat{X}^2_{+} \rangle + \langle \hat{P}^2_{-} \rangle < 1 \label{eq:Duan}
\end{equation} 
is inseparable. Note that this could equally well be formulated in terms of $\hat{X}_-$ and $\hat{P}_+$, though Eq.~(\ref{eq:Duan}) shall be the suitable form here. 

\begin{figure}[th]
\begin{center}
\includegraphics[scale=0.3]{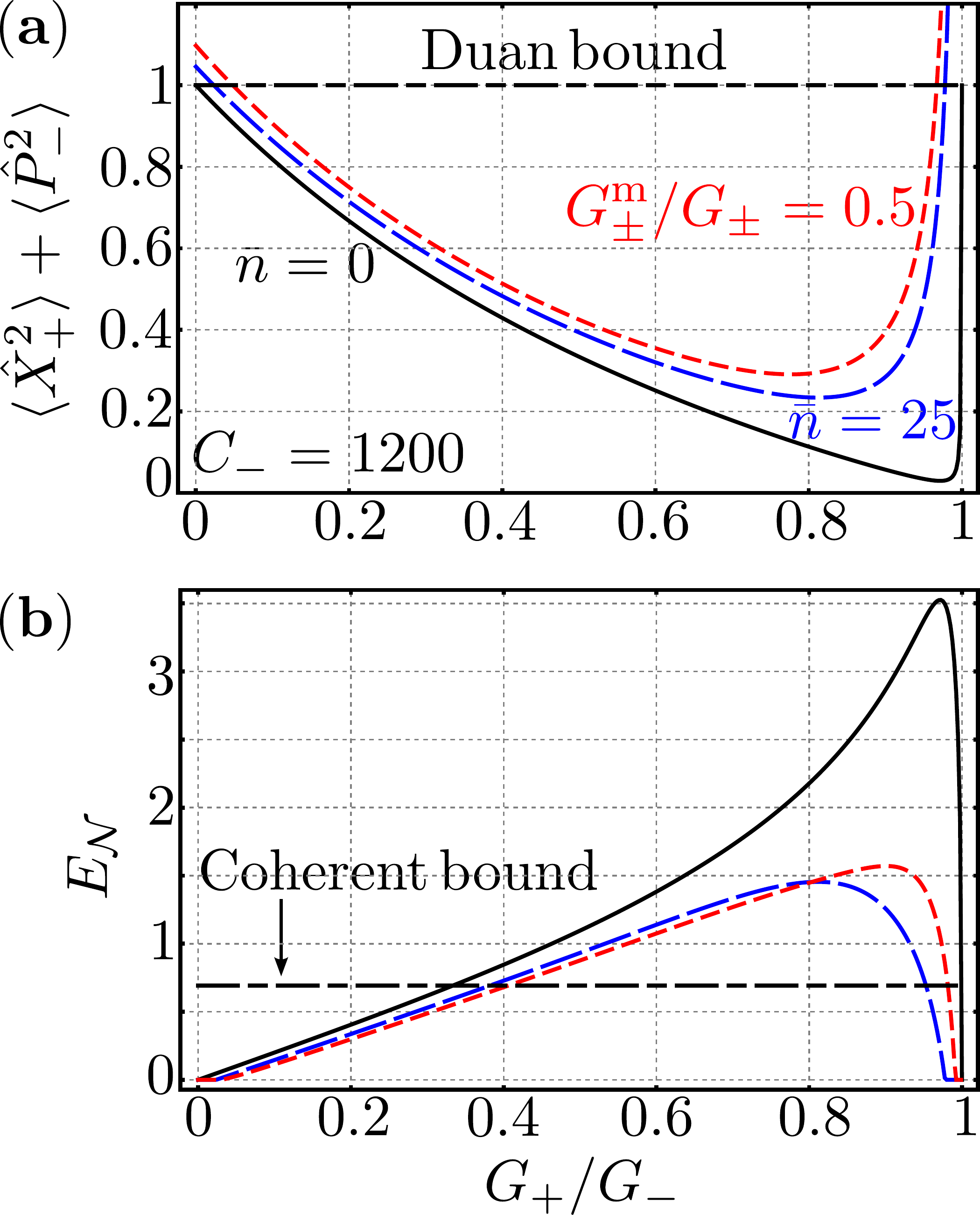} 
\end{center}
\caption{  (Color online) Entanglement, expressed via (a) the Duan quantity (\ref{eq:Duan}) and (b) the logarithmic negativity, against the drive asymmetry, $G_+/G_-$. We hold $G_-$ fixed (fixing the cooperativity $C_-$) and vary $G_+$. These results correspond to the time-independent (rotating-wave approximation) Hamiltonian (\ref{eq:HamGen1}), and apply to both the two-tone and four-tone driving cases, with the effective couplings $G_{\pm}$ given by Eqs.~(\ref{eq:optomechTwo}) and (\ref{eq:GPlusMinus}), respectively. These plots are obtained using the adiabatic limit results of Sec.~\ref{sec:adiabatic}, though they coincide with the results for the full system obtained in Sec.~\ref{sec:fullRWA}. The solid black curve corresponds to a mechanical bath thermal occupation of $\bar{n}=0$, and $G^{\rm m}_{\pm} = 0$ where $G^{\rm m}_{\pm}$ are the effective coupling imperfections, introduced in Eqs.~(\ref{eq:HamBetter}) and (\ref{eq:Gmismatched}) for the two-tone and four-tone driving cases, respectively. The blue curve (long dashes) corresponds to $\bar{n}=25$ and $G^{\rm m}_{\pm} = 0$, while the red curve (short dashes) corresponds to $\bar{n}=25$ and $G^{\rm m}_{\pm} = 0.5G_{\pm}$. Parameters common to each curve are: $C_- = 1200$, $\kappa = 2\pi \times 1.592 \times 10^5 \, {\rm Hz} = 10^6 \, {\rm s^{-1}}$, $\gamma/\kappa = 4 \times 10^{-5}$, and $\Omega/\kappa = 0.1$. }
\label{fig:EntanglementPlots}
\end{figure}

For our system we consider the limit $\gamma/\Omega \rightarrow 0$, since this ensures that the sum and difference Bogoliubov modes are sufficiently coupled (or equivalently, that the two individual Bogoliubov modes see effectively independent reservoirs). For $G_- \neq G_+$, we find for the steady-state second moments:
\begin{subequations}
\begin{eqnarray}
\langle \hat{X}^2_{\pm} \rangle = \langle \hat{P}^2_{\mp} \rangle & = & \frac{\gamma}{\gamma + \Gamma} (\bar{n} + 1/2 ) + \frac{\Gamma}{\gamma + \Gamma} \frac{e^{\mp 2r}}{2} , \nonumber \\
& & \label{eq:collectiveOmegaBig} \\
& = & \frac{\gamma \kappa}{\gamma \kappa + 4 (G^2_- - G^2_+)} (\bar{n} + 1/2) \nonumber \\
& & + \frac{2(G_- \mp G_+)^2}{\gamma \kappa + 4(G^2_- - G^2_+)} . \label{eq:collectiveG+G-} 
\end{eqnarray}
\end{subequations} 
Eq.~(\ref{eq:collectiveOmegaBig}) takes a particularly simple form, describing coupling to a squeezed reservoir with an optomechanical damping rate $\Gamma$. The results (\ref{eq:collectiveOmegaBig}) and (\ref{eq:collectiveG+G-}) are easily checked against the solution of the full system (i.e. without the adiabatic elimination), as discussed in Sec.~\ref{sec:full}. 

The $G_+ = G_- $ limit is unclear from Eq.~(\ref{eq:collectiveOmegaBig}), as it corresponds to the limits $\Gamma \rightarrow 0$ and $r \rightarrow +\infty$. However, the result is clear from Eq.~(\ref{eq:collectiveG+G-}), and we recover the result that $\langle \hat{X}^2_+ \rangle = \langle \hat{P}^2_- \rangle = \bar{n} + 1/2$ (evading the back-action) and $\langle \hat{X}^2_- \rangle = \langle \hat{P}^2_+ \rangle = \bar{n} + 1/2 + C_{\pm}/2$ (heated by the back-action) \cite{woolley:twomode}, where the cooperativities associated with the blue and red sideband drives alone (denoted by the subscripts ``+'' and ``-'', respectively), are
\begin{equation}
C_{\pm} \equiv \frac{\Gamma_{\pm}}{\gamma} \equiv \frac{4G^2_{\pm}}{\gamma \kappa } . \label{eq:cooperativityparameter}
\end{equation}

\begin{figure}[th]
\begin{center}
\includegraphics[scale=0.3]{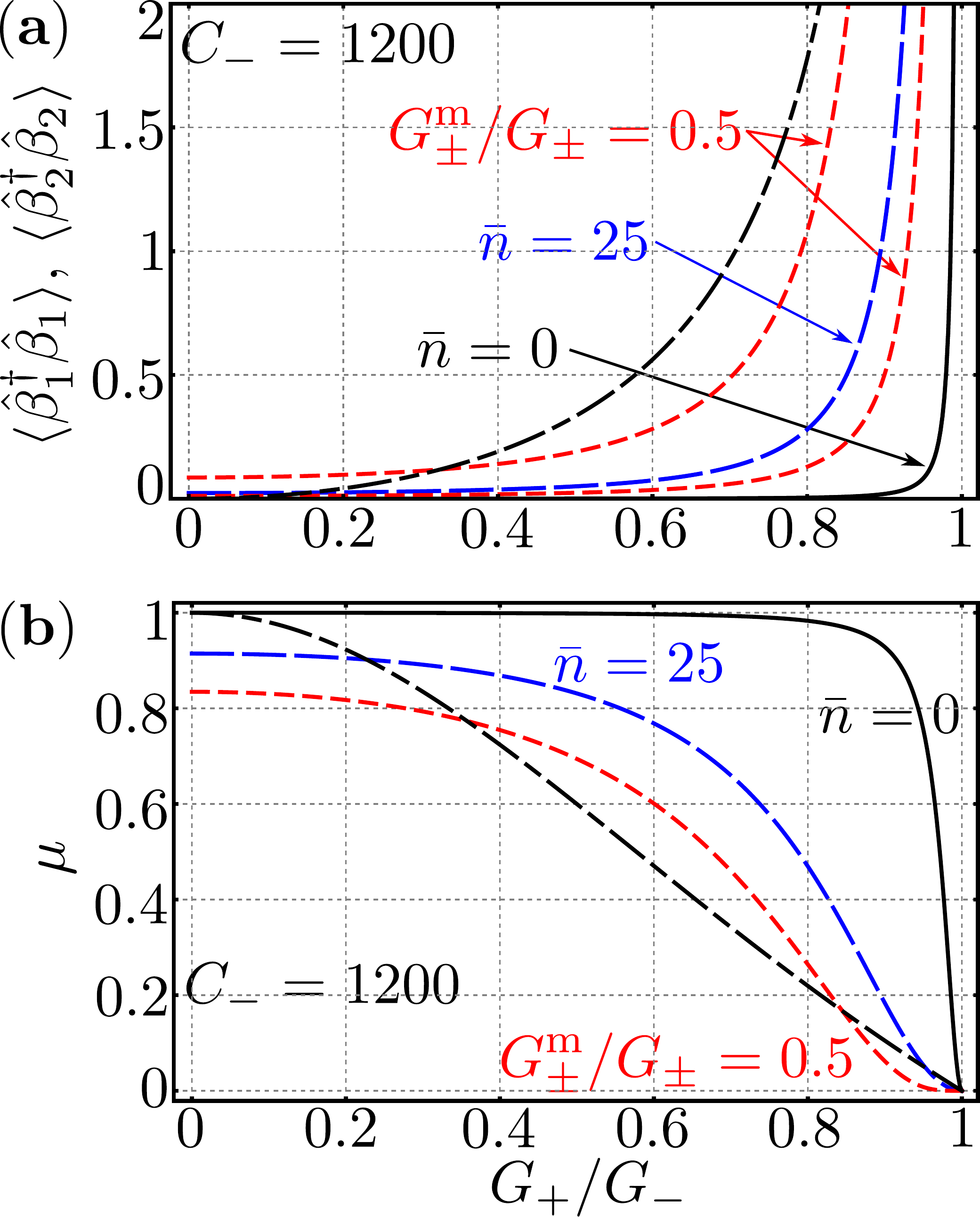}
\end{center}
\caption{ (Color online) (a) Steady-state occupations of the mechanical Bogoliubov modes, defined in Eqs.~(\ref{eq:bog1}) and (\ref{eq:bog2}), and (b) steady-state purity of the two mechanical modes, defined in Eq.~(\ref{eq:purityDef}), against drive asymmetry. The solid black curve corresponds to a mechanical bath thermal occupation of $\bar{n}=0$ and no imperfection in the effective couplings ($G^{\rm m}_{\pm} = 0$), the blue curve (long dashes) corresponds to $\bar{n}=25$ and $G^{\rm m}_{\pm} = 0$, and the red curve (short dashes) corresponds to $\bar{n}=25$ and $G^{\rm m}_{\pm} = 0.5G_{\pm}$. Only one solid black curve and one dashed blue curve is shown in (a) since the occupations of the two Bogoliubov modes are the same in each of these cases. The black curves with long/short dashes correspond to bounds on these quantities for the scheme in which only one Bogoliubov mode is cooled \cite{wang:coolBogoliubov}, assuming a thermal occupation of $\bar{n}=0$: the curve in (a) is a lower bound on the occupation of the uncooled Bogoliubov mode, while the curve in (b) is an upper bound on the purity of the steady-state in this case. Remaining parameters for each curve are as given in the caption of Fig.~\ref{fig:EntanglementPlots}. }
\label{fig:PurityPlots}
\end{figure}

While the Duan inequality provides a simple entanglement criterion, the entanglement may be quantified via the logarithmic negativity, defined in App.~\ref{sec:entanglementandpurity}. Both are shown, as functions of the drive asymmetry $G_+/G_-$, in Fig.~\ref{fig:EntanglementPlots}. In Fig.~\ref{fig:EntanglementPlots}(a), it is seen that the Duan quantity takes a value below one for experimentally reasonable parameters, achievable in state-of-the-art microwave cavity optomechanics experiments \cite{teufel,massel,schwab4}, indicating that the mechanical oscillators are entangled in the steady-state. This continues to be the case even when one accounts for large non-zero initial thermal occupations and large imperfections in the effective couplings. Further, the logarithmic negativity, shown in Fig.~\ref{fig:EntanglementPlots}(b), takes a large value for these parameters. For comparison, the logarithmic negativity of a stationary two-mode squeezed state generated via a Hamiltonian parametric amplifier interaction is bounded above, due to a stability constraint, by $E_{\mathcal{N}} = {\rm ln} \, 2 \sim 0.69$. 

As previously noted \cite{wang:coolBogoliubov,kronwald}, the entanglement goes through a maximum as a function of the drive asymmetry. For the mechanical steady-state to be highly entangled we require both that the target steady-state is highly squeezed ($r$ large, requiring $G_+/G_- \rightarrow 1$) \emph{and} that the system is effectively cooled towards this steady-state ($\Gamma$ large, requiring $G_+/G_- \rightarrow 0$). Obviously, these limits are incompatible and the optimal asymmetry is between these, leading to the observed maximum.  

Given the simple analytical results we have obtained, we may optimise the steady-state entanglement analytically by minimizing the Duan quantity over the drive asymmetry. This is most conveniently done using Eq.~(\ref{eq:collectiveG+G-}). We find that the optimal drive asymmetry is
\begin{subequations}
\begin{eqnarray}
\left. \frac{G_+}{G_-} \right|_{\rm opt.} & = & 1 + \frac{1+\bar{n}}{C_-} - \sqrt{ \frac{1+1/C_-}{C_-} }  \label{eq:driveasym} \\
& \sim & 1 - \frac{1}{ \sqrt{C_-} } . \label{eq:largeC}
\end{eqnarray} 
\end{subequations}
The result (\ref{eq:largeC}) holds in the large-cooperativity limit, provided that one is still within the adiabatic regime. It follows that the Duan quantity (\ref{eq:Duan}), to first-order in $C^{-1}_-$, is
\begin{equation}
\langle \hat{X}^2_+\rangle + \langle \hat{P}^2_- \rangle = \frac{1+\bar{n}}{ \sqrt{ C_- }} + \frac{\bar{n} (1+\bar{n})}{C_-} .
\end{equation}
Clearly, the mechanical oscillators are entangled even for a modest cooperativity. We emphasize these results are only valid in the adiabatic limit, corresponding to $C_- \leq 4\kappa/\gamma$. The achievable entanglement beyond the adiabatic regime shall be discussed in Sec.~\ref{sec:full}. 

\subsection{Purity}
The purity of the steady-state generated is relevant for both experiments in quantum foundations and in quantum information processing; its role in determining teleportation fidelity shall be described in Sec.~\ref{sec:teleportationfidelity}. Now, the fact that the steady-state is highly-entangled does not necessarily imply that the steady-state is highly-pure. Indeed, if one cools only one Bogoliubov mode then the steady-state is highly-entangled, but also highly impure \cite{wang:coolBogoliubov}. The purity of the mechanical two-mode steady-state is defined as 
\begin{equation}
\mu \equiv {\rm tr} (\rho^2), \label{eq:purityDef}
\end{equation}
where $\rho$ is the density matrix of the two mechanical modes. It can be directly evaluated from knowledge of its symmetrically-ordered covariance matrix $\mathbf{V}$. With quadratures as defined in Eqs.~(\ref{eq:usualquadratures}) and the covariance matrix expressed in the ordered basis $( \hat{X}_a,\hat{P}_a,\hat{X}_b,\hat{P}_b)$, the purity is simply given by 
\begin{equation}
\mu = \left. 1 \middle/ \left( 4 \sqrt{ \det \mathbf{V} } \right) \right. .
\end{equation}
The purity may also be assessed by calculating the thermal occupations of the two mechanical Bogoliubov modes, defined in Eqs.~(\ref{eq:bog1}) and (\ref{eq:bog2}). 

Plots of both quantities, against the drive asymmetry $G_+/G_-$, are shown in Fig.~\ref{fig:PurityPlots}. From Fig.~\ref{fig:PurityPlots}(a) it is clear that the occupations of the two Bogoliubov modes are the same provided that the imperfections in the effective couplings, Eqs.~(\ref{eq:coupleimperfect2}) and (\ref{eq:Gmismatched}), are zero. Further, the occupations are close to zero for reasonable parameters, verifying that our scheme effectively cools both Bogoliubov modes. The purity of the state is correspondingly high, being close to one for reasonable experimental parameters, see Fig.~\ref{fig:PurityPlots}(b), and so out-performing a scheme in which only one Bogoliubov mode is cooled \cite{wang:coolBogoliubov}. 

In the absence of effective coupling imperfections, we can obtain simple analytical results characterizing the purity of the steady-state. The occupations of the Bogoliubov modes in the limit $\gamma/\Omega \rightarrow 0$ (and for $G_- \neq G_+$) are
\begin{subequations}
\begin{eqnarray}
\langle \hat{\beta}^\dagger_i \hat{\beta}_i \rangle & = & \frac{\gamma}{\gamma + \Gamma} \left[ \bar{n} + (2\bar{n} + 1 ) \sinh^2 r \right] \\
& = & \frac{\gamma \kappa}{\gamma \kappa + 4 ( G^2_- - G^2_+ ) } \left[ \frac{ G^2_+ + \bar{n} ( G^2_- + G^2_+ ) }{G^2_- - G^2_+} \right] , \nonumber \\
& & 
\end{eqnarray}
\end{subequations}
for $i=1$ and $2$, consistent with the result for single-mode squeezing in Ref.~\onlinecite{kronwald}. Again, it is clear that both Bogoliubov modes are cooled equally. The purity itself, in the limit $\gamma/\Omega \rightarrow 0$, is given by 
\begin{eqnarray}
\mu & = & \frac{ \left( \gamma + \Gamma \right)^2 }{ \left[ \gamma (1 + 2\bar{n}) + \Gamma \right]^2 + 4 (1 + 2\bar{n}) \gamma \Gamma \sinh^2 r } . \label{eq:purityexplicit} \nonumber \\
& & 
\end{eqnarray}
The purity at the entanglement maximum, in the large-cooperativity limit, is $\mu \sim 1/[2( 1 + \bar{n} ) ] $. However, a highly-pure (and still highly-entangled) steady-state can be achieved by choosing a drive asymmetry just below that corresponding to the optimal entanglement (as this gives a large $\Gamma$ and hence more cooling, at the expense of a smaller squeeze parameter $r$). 

\begin{figure}[ht]
\begin{center}
\includegraphics[scale=0.3]{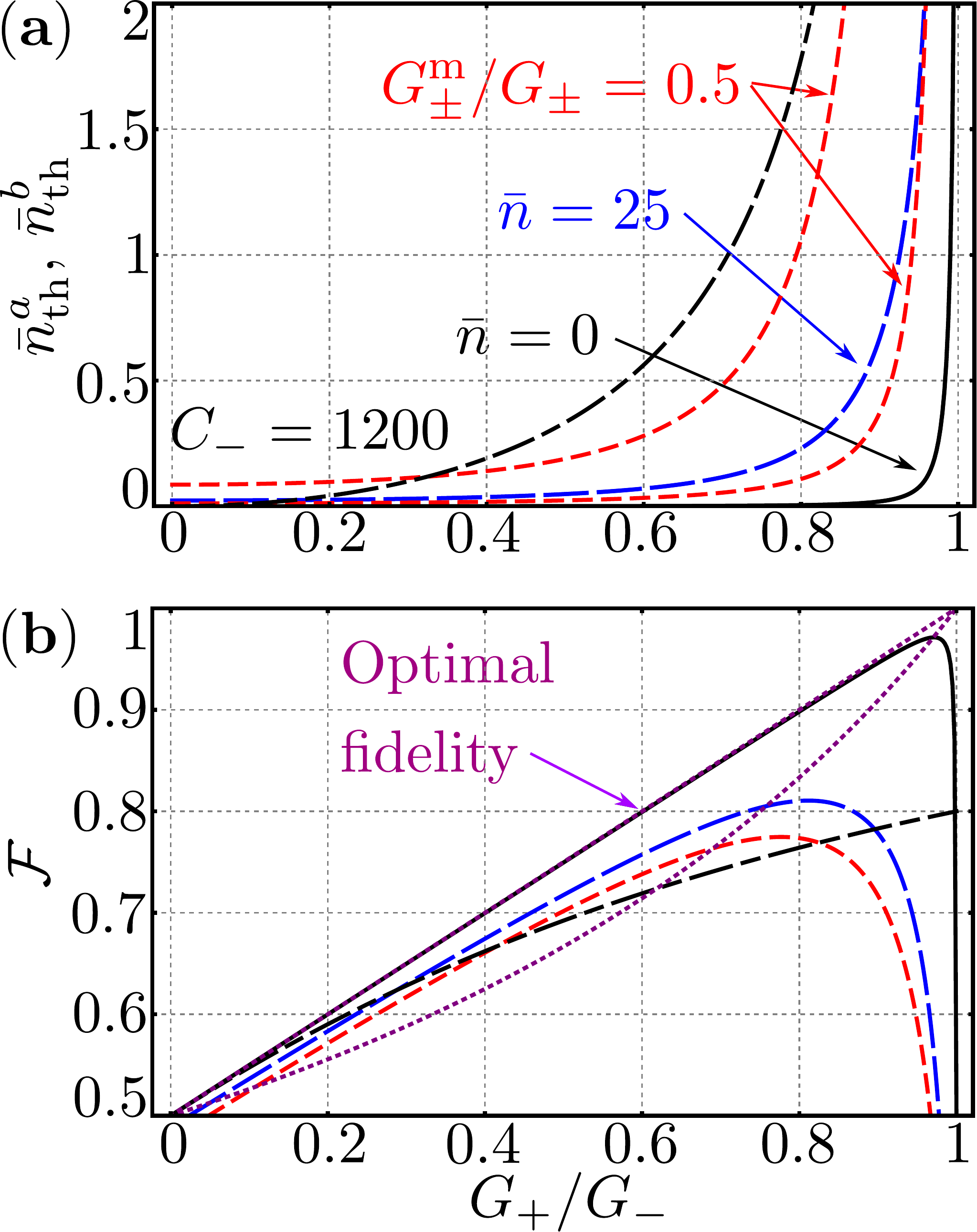} 
\end{center}
\caption{ (Color online) (a) Effective thermal occupations for each mechanical mode in the steady-state, introduced in Eq.~(\ref{eq:thermalTMSSmain}), against the drive asymmetry. (b) The teleportation fidelity, using the generated mechanical steady-state as an EPR channel to teleport a coherent state via the standard protocol, against the drive asymmetry. The solid black curves correspond to a mechanical bath thermal occupation of $\bar{n} = 0$ and no imperfection in the effective couplings ($G^{\rm m}_{\pm} = 0$), the blue curves (long dashes) correspond to $\bar{n} = 25$ and $G^{\rm m}_{\pm} = 0$, and the red curves (short dashes) correspond to $\bar{n} = 25$ and $G^{\rm m}_{\pm} = 0.5G_{\pm}$. The black curve with long/short dashes on the left is the lower bound on the occupation of the uncoupled mode in a scheme that cools one Bogoliubov mode \cite{wang:coolBogoliubov}. The black curve with long/short dashes on the right is the teleportation fidelity achievable with such a scheme. The upper (lower) dotted purple curve is the upper (lower) bound on the optimal teleportation fidelity achievable for a given amount of entanglement. The given amount of entanglement corresponds to that possessed by a two-mode squeezed state with squeezing parameter $r = \rm{tanh}^{-1} \, (G_+/G_-)$, c.f. Eq.~(\ref{eq:r}). Remaining parameters for each solid curve are as given in the caption of Fig.~\ref{fig:EntanglementPlots}. }
\label{fig:TeleportationPlots}
\end{figure}

\subsection{Teleportation fidelity}
\label{sec:teleportationfidelity}
The two-mode squeezed state generated here may be regarded as the entangled resource (``EPR channel'') in a continuous-variable teleportation protocol \cite{braunstein}. If we write the steady-state two-mode symmetrically-ordered covariance matrix, in the ordered basis $( \hat{X}_a, \hat{P}_a, \hat{X}_b, \hat{P}_b )$, in block form as
\begin{equation}
\mathbf{V} = \left[ \begin{array}{c|c} \mathbf{V}_a & \mathbf{V}_{ab} \\ \hline \mathbf{V}^T_{ab} & \mathbf{V}_b \end{array} \right] , \label{eq:blockform}
\end{equation}  
then the teleportation fidelity for a single-mode Gaussian input state under the standard protocol is given by \cite{fiurasek}
\begin{subequations}
\begin{eqnarray}
\mathcal{F} & = & \frac{1}{ \sqrt{\det (2\mathbf{V}_{\rm in} + \mathbf{N}) } } , \\
\mathbf{N} & = & \sigma_z \mathbf{V}_a \sigma_z + \sigma_z \mathbf{V}_{ab} + \mathbf{V}^T_{ab} \sigma_z + \mathbf{V}_b ,
\end{eqnarray}
\end{subequations}
where $\mathbf{V}_{\rm in}$ is the covariance matrix of the state to be teleported. For the teleportation of a coherent state this is $\mathbf{V}_{\rm in} = (1/2) \mathbf{I}_2$. 

We find that the mechanical steady-state in our system, in the limit $\gamma/\Omega \rightarrow 0$, is a \emph{thermal} two-mode squeezed state. Such a state is defined by
\begin{eqnarray}
\rho & \equiv & \hat{S}_{2} (\xi ) \left( \rho^a_{\rm th} \otimes \rho^b_{\rm th} \right) \hat{S}^\dagger_{2} (\xi ) , \label{eq:thermalTMSSmain}
\end{eqnarray}
where $\rho^{a(b)}_{\rm th}$ denotes the density matrix of a thermal state of mode $a(b)$ with occupation $\bar{n}^{a(b)}_{\rm th}$, and $\hat{S}_{2}(\xi )$ is the two-mode squeezing operator introduced in Eq.~(\ref{eq:TMSoperator}). Accordingly, we can assign effective occupations, $\bar{n}^{a(b)}_{\rm th}$, and an effective two-mode squeezing parameter, $\xi$, to our steady-state, as detailed in App.~\ref{sec:TTMSS}. Now the purity of such a state is simply 
\begin{equation}
\mu = \frac{1}{ (1+2\bar{n}^a_{\rm th}) (1+2\bar{n}^b_{\rm th}) } ,
\end{equation}
and the teleportation fidelity, based on a thermal two-mode squeezed state channel, is \cite{fiurasek}
\begin{equation}
\mathcal{F} = \frac{1}{e^{-2\xi} ( 1 + \bar{n}^a_{\rm th} + \bar{n}^b_{\rm th} + e^{2\xi}) } . \label{eq:teleportationfidelity}
\end{equation}
Clearly, larger effective occupations correspond to a lower purity and a lower teleportation fidelity.

It is known that when the channel of a continuous-variable teleportation protocol is a \emph{symmetric} thermal two-mode squeezed state ($\bar{n}^a_{\rm th}, \bar{n}^b_{\rm th} \equiv \bar{n}_{\rm th}$), the teleportation fidelity is simply \cite{adessoilluminati}
\begin{equation}
\mathcal{F} = \frac{1}{1 + e^{-E_{\mathcal{N}}}} , \label{eq:simplefidelity}
\end{equation}
where the logarithmic negativity is given by $ E_{\mathcal{N}} = {\rm Max} \left[ 0 , 2\xi - {\rm ln} ( 1 + 2\bar{n}_{\rm th} ) \right] $. Eq.~(\ref{eq:simplefidelity}) actually gives the optimal teleportation fidelity achievable for a given amount of entanglement \cite{adessoilluminati}. Asymmetry results in a teleportation fidelity below that given by Eq.~(\ref{eq:simplefidelity}). 

The effective thermal occupations for our steady-state ($\bar{n}^a_{\rm th}$ and $\bar{n}^b_{\rm th}$), as a function of the drive asymmetry, are shown in Fig.~\ref{fig:TeleportationPlots}(a). It is clear that $\bar{n}^a_{\rm th} = \bar{n}^b_{\rm th}$ provided that the imperfections in the effective couplings are zero. Accordingly in these cases, the teleportation fidelity is given by Eq.~(\ref{eq:simplefidelity}) and so attains its optimal value for a given amount of entanglement. 

Consider now the case of an asymmetric channel, $\bar{n}^{\rm th}_a \neq \bar{n}^{\rm th}_b$; this is generically the kind of state produced using a scheme which cools only a single Bogoliubov mode \cite{wang:coolBogoliubov}. For such states the standard teleportation protocol does not achieve the fidelity in Eq.~(\ref{eq:simplefidelity}). This fidelity can be reached in principle in the highly-entangled regime, if one goes beyond the standard protocol by allowing for additional local Gaussian operations \cite{mari}. 

The teleportation fidelity, assuming that the mechanical steady-state we have generated is used as an EPR channel and assuming no other sources of imperfection, is plotted as a function of drive asymmetry in Fig.~\ref{fig:TeleportationPlots}(b). Crucially, by cooling both Bogoliubov modes the optimal teleportation fidelity tends to $1$ rather than to $4/5$ in the highly-entangled regime. Further, this upper bound is achievable with reasonable parameters. 
  
\subsection{Coherent feedback }
\label{sec:coherentfeedback}
\begin{figure}[ht]
\begin{center}
\includegraphics[scale=0.46]{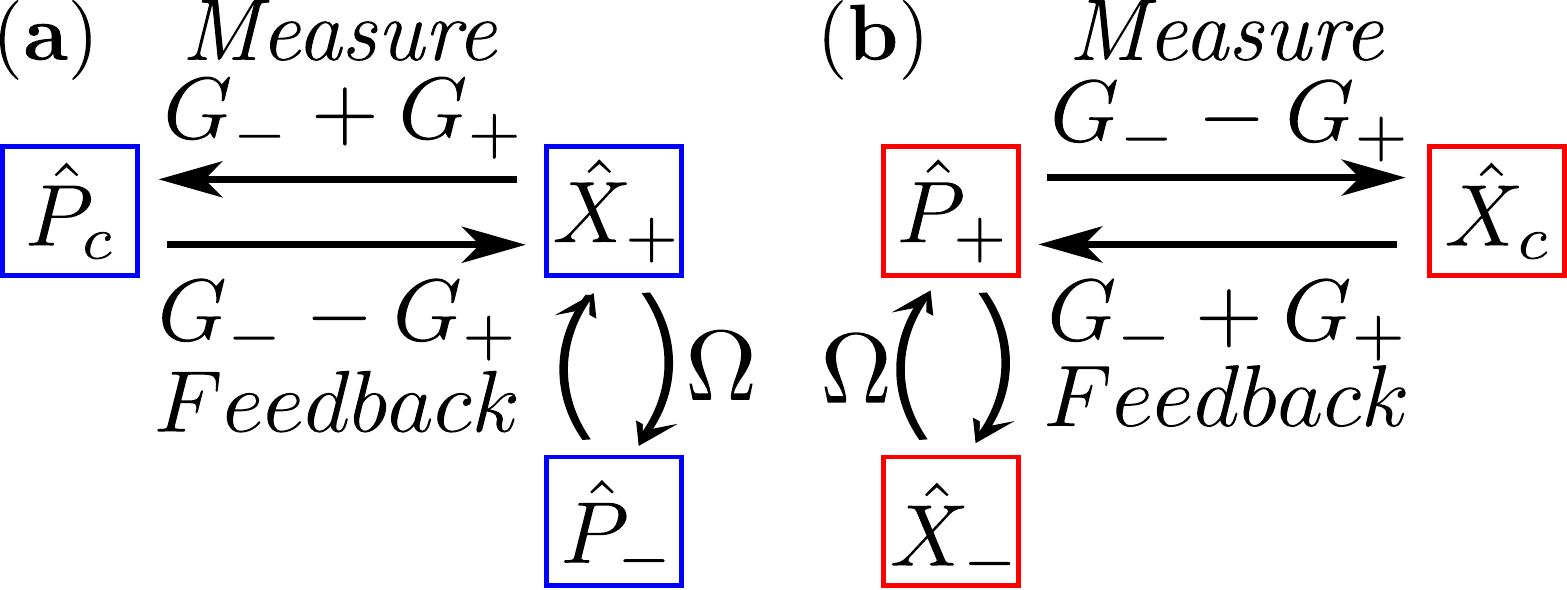}
\end{center}
\caption{  (Color online) Representation of the reservoir engineering scheme of Figs.~\ref{fig:optodualmechanics} and \ref{fig:optomechanics4tone} as a form of coherent feedback, with the feedback (back-action) being applied autonomously via the cavity mode. The description is in terms of the collective mechanical quadrature operators introduced in Eqs.~(\ref{eq:collectivequad1}) and (\ref{eq:collectivequad2}). (a) There is a measurement of $\hat{X}_+$ via the cavity at a rate $\sim (G_- + G_+)$ and a feedback onto $\hat{X}_+$ via the cavity at a rate $\sim (G_- - G_+)$. The measurement rate being greater than the feedback rate leads to squeezing. (b) There is a measurement of $\hat{P}_+$ via the cavity at a rate $\sim (G_- - G_+)$ and a feedback onto $\hat{P}_+$ via the cavity at a rate $\sim (G_- + G_+)$. The feedback rate being greater than the measurement rate leads to amplification. } 
\label{fig:coherentSchematicEdit}
\end{figure}

The reservoir engineering scheme that we have described here permits an alternative interpretation in terms of coherent feedback \cite{kerckhoff}, similar to that provided for the squeezing scheme in Ref.~\onlinecite{kronwald}; we depict the process schematically in Fig.~\ref{fig:coherentSchematicEdit}. The Hamiltonian (\ref{eq:HamGen1}) can be rewritten in terms of the collective mechanical quadratures of Eqs.~(\ref{eq:collectivequad1}) and (\ref{eq:collectivequad2}) as
\begin{eqnarray}
\hat{\mathcal{H}} & = & \Omega \left( \hat{X}_+ \hat{X}_- + \hat{P}_+ \hat{P}_- \right) + \sqrt{2} (G_- + G_+) \hat{X}_+ \hat{X}_c \nonumber \\
& & + \sqrt{2} (G_- - G_+) \hat{P}_+ \hat{P}_c + \hat{H}_{\rm diss} . \label{eq:cohFB}
\end{eqnarray}
This Hamiltonian is a perturbation, via the third term, of a Hamiltonian that we previously studied in the context of two-mode back-action-evading measurement and feedback control \cite{woolley:twomode}. From a \emph{coherent} feedback point of view, the second term in Eq.~(\ref{eq:cohFB}) may be regarded as the ``measurement'' interaction and the third term in Eq.~(\ref{eq:cohFB}) may be regarded as the ``feedback'' back-action, applied autonomously via the cavity mode. 

After an adiabatic elimination of the cavity mode, the Heisenberg-Langevin equations corresponding to the Hamiltonian (\ref{eq:cohFB}) may be written as 
\begin{equation}
\frac{d}{dt}\vec{X} = \mathbf{A}_0 \cdot \vec{X} + \mathbf{B}_1 \cdot \vec{X}_{\rm in} + \mathbf{B}_2 \cdot \vec{Y}_{\rm in}, \label{eq:adiabaticlangevin}
\end{equation}
where $\vec{X} = ( \hat{X}_+, \hat{P}_+, \hat{X}_-, \hat{P}_- )^T$ is the vector of mechanical collective quadrature observables, $\vec{X}_{\rm in}$ is the corresponding vector of mechanical input noises, and $\vec{Y}_{\rm in} \equiv [ \hat{Y}_1(t), \hat{Y}_2 (t) ]^T$ are the operators associated with the cavity input noise. The remaining matrices in Eq.~(\ref{eq:adiabaticlangevin}) are specified in App.~\ref{sec:HLcoll}. The new noise input operators have the correlation functions 
\begin{subequations}
\begin{eqnarray}
\langle \hat{Y}_{1}(t) \hat{Y}_{1} ( 0 ) \rangle = \frac{1}{2} \frac{G_- - G_+}{G_- + G_+} \delta (t) \equiv \left(\bar{n}_1 + \frac{1}{2} \right) \delta (t) , \nonumber \\ 
& & \\
\langle \hat{Y}_{2}(t) \hat{Y}_{2} ( 0 ) \rangle = \frac{1}{2} \frac{G_- + G_+}{G_- - G_+} \delta (t) \equiv \left( \bar{n}_2 + \frac{1}{2} \right) \delta (t) , \nonumber \\
& & 
\end{eqnarray}
\end{subequations}
where $\bar{n}_1$ and $\bar{n}_2$ denote effective thermal occupations of the noise inputs. Clearly, the input noises seen by the collective mechanical quadratures are weighted by the ratios of the measurement and feedback rates. Since we have $G_- > G_+ \geq 0$ here, the effective occupation $\bar{n}_1$ is negative \cite{kronwald}. Therefore, as far as the collective quadrature $\hat{X}_+$ is concerned, the cavity behaves as a squeezed bath and therefore $\bar{X}_+$ will be squeezed in the steady-state \cite{walls}. Conversely, $\bar{n}_2$ is positive and the collective quadrature $\hat{X}_-$ will be anti-squeezed in the steady-state. 

\section{Full System}
\label{sec:full}

\begin{figure*}[ht]
\begin{center}
\includegraphics[scale=0.5]{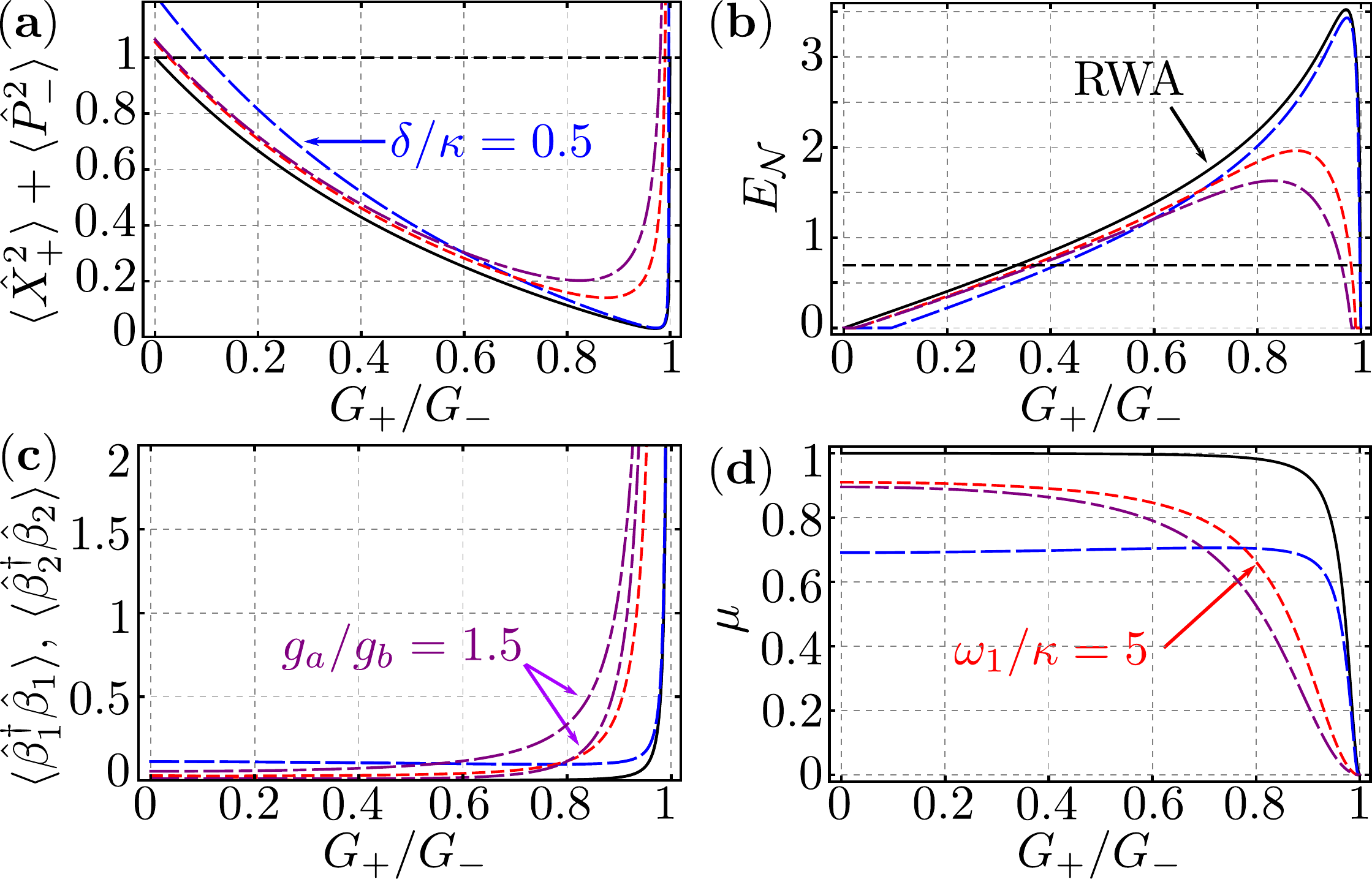}
\end{center}
\caption{  (Color online) The effects of counter-rotating contributions to the Hamiltonian, described in detail in App.~\ref{sec:CR}, on the entanglement and purity of the mechanical two-mode steady-state under four-tone driving. The plots show (a) the Duan quantity, (b) the logarithmic negativity, (c) the occupations of the Bogoliubov modes, and (d) the purity, as functions of the drive asymmetry. The solid black curve corresponds to the time-independent Hamiltonian (\ref{eq:HamGen1}), while the other curves correspond to cases where counter-rotating effects are significant. The blue curve (long dashes) is for $\delta /\kappa = 0.5$, the red curve (short dashes) is for $\omega_1/\kappa = 5$, and the purple curve (long/short dashes) is for $g_a/g_b=1.5$. Although it is possible for counter-rotating contributions to have a significant effect on these measures, it should also be possible to achieve sufficiently high sideband resolution that their effects may be neglected. Note that under two-tone driving we typically have $g_a = g_b$ and always have $\delta = 0$; the effect of the parameter $\omega_m/\kappa$ in that case is comparable to the effect of the parameter $\omega_1/\kappa$ in the four-tone driving case. Parameters, unless otherwise specified, for each curve are $\delta /\kappa = 1$, $\omega_1/\kappa = 100$ and $d=1$. The parameters common to each curve are $\bar{n}=0$, $G^{\rm m}_{\pm} = 0$, $C_- = 1200$, $\kappa = 2\pi \times 1.592 \times 10^5 \, {\rm Hz} = 10^6 \, {\rm s^{-1}}$, $\gamma/\kappa = 4 \times 10^{-5}$, and $\Omega/\kappa = 0.1$. } 
\label{fig:CR}
\end{figure*}

\subsection{Solution with time-independent Hamiltonian}
\label{sec:fullRWA}
In Sec.~\ref{sec:adiabatic} we calculated the steady-state of our system in the adiabatic limit, after mathematically removing the cavity mode from the system. Even with the cavity mode retained, the effective Hamiltonian (\ref{eq:HamGen1}) is quadratic and time-independent, and we may readily solve for the steady-state. The Heisenberg-Langevin equations corresponding to (\ref{eq:HamGen1}) may be written as the system 
\begin{equation}
\frac{d}{dt} \vec{X} = \mathbf{A}_0 \cdot \vec{X} + \mathbf{B}_0 \cdot \vec{X}_{\rm in} , \label{eq:HLeqns}
\end{equation}
where $\vec{X} = ( \hat{X}_a, \hat{P}_a, \hat{X}_b, \hat{P}_b, \hat{X}_c, \hat{P}_c )^T$ is defined in terms of individual oscillator quadratures, and the matrices are specified in App.~\ref{sec:HLind}. The steady-state, symmetrically-ordered covariance matrix $\mathbf{V}$ is obtained by solving the Lyapunov equation, 
\begin{equation}
\mathbf{A}_0 \mathbf{V} + \mathbf{V} \mathbf{A}^T_0 = -\mathbf{B}_0 \mathbf{B}^T_0. \label{eq:Lyapunov}
\end{equation} 
Solving Eq.~(\ref{eq:Lyapunov}) allows us to assess the steady-state even when we are not in the adiabatic limit, and to validate results obtained in the adiabatic limit. As before, knowledge of the covariance matrix allows the evaluation of entanglement, purity and fidelity measures. The analytical results are easily obtained but sufficiently complicated that we do not quote them here, while the numerical results coincide with those previously obtained in the adiabatic limit.

\subsection{Solution with time-dependent Hamiltonian}
The results of Sec.~\ref{sec:fullRWA} are still only valid provided that we are justified in making a rotating-wave approximation; that is, in discarding the time-dependent (``counter-rotating'') contributions that arise in the derivation of the Hamiltonian (\ref{eq:HamGen1}). Here we account for these time-dependent contributions; the exact forms that they take are given in App.~\ref{sec:HamCR}. The corresponding contributions to the Heisenberg-Langevin equations can be handled by making the replacement $\mathbf{A}_0 \rightarrow \mathbf{A}(t)$ in Eq.~(\ref{eq:HLeqns}), where the time-dependent drift matrix is given by
\begin{equation}
\mathbf{A}(t) = \mathbf{A}_0 + \sum^N_{k=1} \left( \mathbf{A}_{k+} e^{+2i\delta_k t} + \mathbf{A}_{k-} e^{-2i\delta_k t} \right) . \label{eq:driftt}
\end{equation}
For the case of two-tone driving (\ref{eq:twotone}), $N=1$ and $\delta_1 = \omega_m$. For four-tone driving (\ref{eq:fourdrives}) we have $N=4$ and $\delta_1 = \delta$, $\delta_2 = \omega_b + \Omega $, $\delta_3 = \omega_m$, and $\delta_4 = \omega_a-\Omega$; see Fig.~\ref{fig:optomechanics4tone} and recall that $\delta$ was introduced in Eq.~(\ref{eq:deltaDef}). The first of these contributions is due to the second drive on the same side of the cavity resonance frequency, while the remaining contributions are due to the drives on the other side of the cavity resonance frequency. The matrices $\mathbf{A}_{k\pm}$ are given in App.~\ref{sec:MatricesCR}. 

Given that the drift matrix (\ref{eq:driftt}) is now time-varying, the covariance matrix $\mathbf{V}(t)$ is given by solution of the Lyapunov-like differential equation, 
\begin{equation}
\dot{\mathbf{V}} = \mathbf{A} \mathbf{V} + \mathbf{V} \mathbf{A}^\dagger + \mathbf{B}_0 \mathbf{B}^T_0 . \label{eq:LyapunovDE}
\end{equation}
The covariance matrix will be oscillatory in the long-time limit; we seek the dc component of its solution. The direct numerical solution of Eq.~(\ref{eq:LyapunovDE}) is inefficient, so instead we use an ansatz to obtain an approximate numerical solution \cite{mari:nonRWA}. The procedure used is outlined in App.~\ref{sec:SolnCR}.

\begin{figure}[ht]
\begin{center}
\includegraphics[scale=0.38]{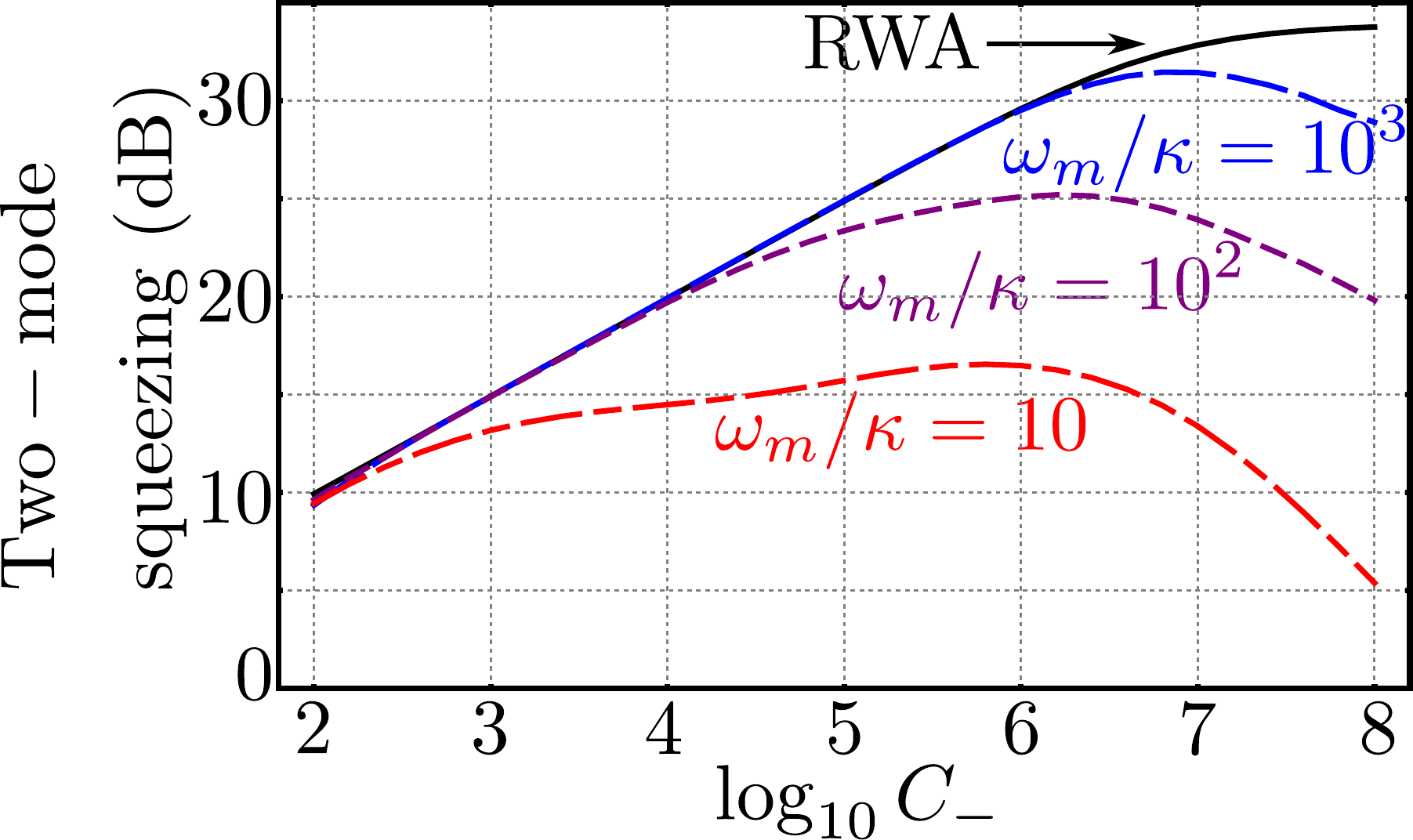}
\end{center}
\caption{  (Color online) Mechanical two-mode squeezing as a function of the cooperativity parameter $C_-$, introduced in Eq.~(\ref{eq:cooperativityparameter}), for a range of sideband resolutions $\omega_m/\kappa$. The results are presented for the case of two-tone cavity driving with the single-photon optomechanical coupling rates being equal, $g_a = g_b$ (that is, no imperfection in the effective coupling rates). The quantity plotted is the two-mode squeezing in dB, defined by $\rm{TMS \ (dB)} \equiv -\log_{10} \left[ \left( \langle \hat{X}^2_+ \rangle + \langle \hat{P}^2_- \rangle \right) / \left(  \langle \hat{X}^2_+ \rangle + \langle \hat{P}^2_- \rangle \right)_0 \right]$, with both the numerator and denominator being instances of the Duan quantity of Eq.~(\ref{eq:Duan}). At each value of the cooperativity parameter $C_-$, the Duan quantity is minimised over the effective coupling asymmetry $G_+/G_-$. The curves are shown for: a rotating-wave approximation (solid black curve) meaning that the sideband resolution is effectively infinite; $\omega_m/\kappa = 10^3$ (blue curve with long dashes); $\omega_m/\kappa = 10^2$ (purple curve with short dashes); $\omega_m/\kappa = 10$ (red curve with long/short dashes). In the case of four-tone cavity driving the behaviour of the two-mode squeezing as a function of $\omega_1/\kappa$ is similar to the behaviour seen here as a function of $\omega_m/\kappa$. The other parameters are as specified in the caption of Fig.~\ref{fig:CR}. } 
\label{fig:CRCooperativity}
\end{figure}

\subsection{Effects of counter-rotating terms}
\label{sec:CRterms}
The effects of the counter-rotating Hamiltonian contributions on the entanglement and purity of the steady-state are shown in Figs.~\ref{fig:CR} and \ref{fig:CRCooperativity}. Fig.~\ref{fig:CR} shows these as functions of drive asymmetry, while Fig.~\ref{fig:CRCooperativity} shows the two-mode squeezing (entanglement), optimized over the drive asymmetry, as a function of the cooperativity parameter $C_-$. 

From Fig.~\ref{fig:CR} it is clear that there is a degradation in the entanglement and purity measures as the frequency of the counter-rotating terms is lowered. This is unsurprising since the form of the time-dependent Hamiltonian contributions, detailed in App.~\ref{sec:HamCR}, depart from the ideal form of Eq.~(\ref{eq:HamCoolAsBogoliubov}). However, the same overall behaviour in the entanglement and purity is observed; that is, a maximum in the entanglement and a monotonic decrease in the purity. 

The rotating-wave approximation results coincide with the full time-dependent Hamiltonian results in the limit that all counter-rotating frequencies greatly exceed the cavity decay rate; that is, all $| \delta_{k} | \gg \kappa$. With the parameters chosen, counter-rotating effects become significant at $\delta/\kappa \sim 0.5$ and $\omega_a/\kappa \sim 5$; these correspond to modest sideband resolutions. It should be possible to significantly exceed this resolution and therefore largely avoid the effects of counter-rotating terms. The ratio $\delta/\kappa$ can be reduced further than the ratio $\omega_a/\kappa$ without significant deleterious effects due to the distinct manner with which the corresponding contributions enter the full time-dependent Hamiltonian. Provided that the asymmetry in the single-photon optomechanical coupling rates is small, the effective couplings associated with the terms rotating at $\pm 2\delta$ are not exponentially enhanced in the large-$r$ limit, while those oscillating at $\pm 2\omega_a$, $\pm 2\omega_m$ and $\pm 2\omega_b$ are exponentially enhanced; see Eq.~(\ref{eq:CRfourBog}).  

Note that with counter-rotating terms included, the entanglement and purity of the mechanical steady-state depend on the asymmetry in the single-photon optomechanical coupling rates, \emph{even if there are no imperfections in the effective couplings}. This is in contrast to the results obtained with the time-independent Hamiltonian (\ref{eq:HamGen1}). As seen in Fig.~\ref{fig:CR}, this asymmetry leads to a significant degradation in entanglement and purity measures if the imperfection is around $\sim$~$50\%$ of the effective coupling. Again, it should be possible to engineer the optomechanical system such that the asymmetry is much lower than this value, and the corresponding deleterious effects are negligible.  

The behaviour of the optimised mechanical two-mode squeezing as a function of the cooperativity parameter $C_-$ is shown in Fig.~\ref{fig:CRCooperativity}. Even neglecting the effects of counter-rotating terms, the achievable two-mode squeezing (entanglement) plateaus in the large-cooperativity limit. For a finite sideband resolution, however, the squeezing goes through a maximum as a function of the cooperativity, with the maximum occurring at a lower value of the cooperativity parameter as the sideband resolution is decreased. Unsurprisingly, the discrepancy between the RWA result and the full result increases as the sideband resolution is reduced. Note, however, that these discrepancies become significant only at \emph{very} large values of the cooperativity parameter. Also note that in the large-cooperativity limit it is possible for the dynamics associated with the full time-dependent Hamiltonian to be unstable where the dynamics associated with the corresponding time-independent Hamiltonian are stable. However, this does not tend to be the case at the optimal drive asymmetry. This potential for the onset of an instability occurs at higher values of the cooperativity than we have previously considered in this work. Our results show that for high levels of steady-state entanglement, the reservoir engineering scheme discussed here is robust against realistic levels of counter-rotating corrections. 

\section{Experimental observability}
\label{sec:expt}

\begin{figure*}[ht]
\begin{center}
\includegraphics[scale=0.51]{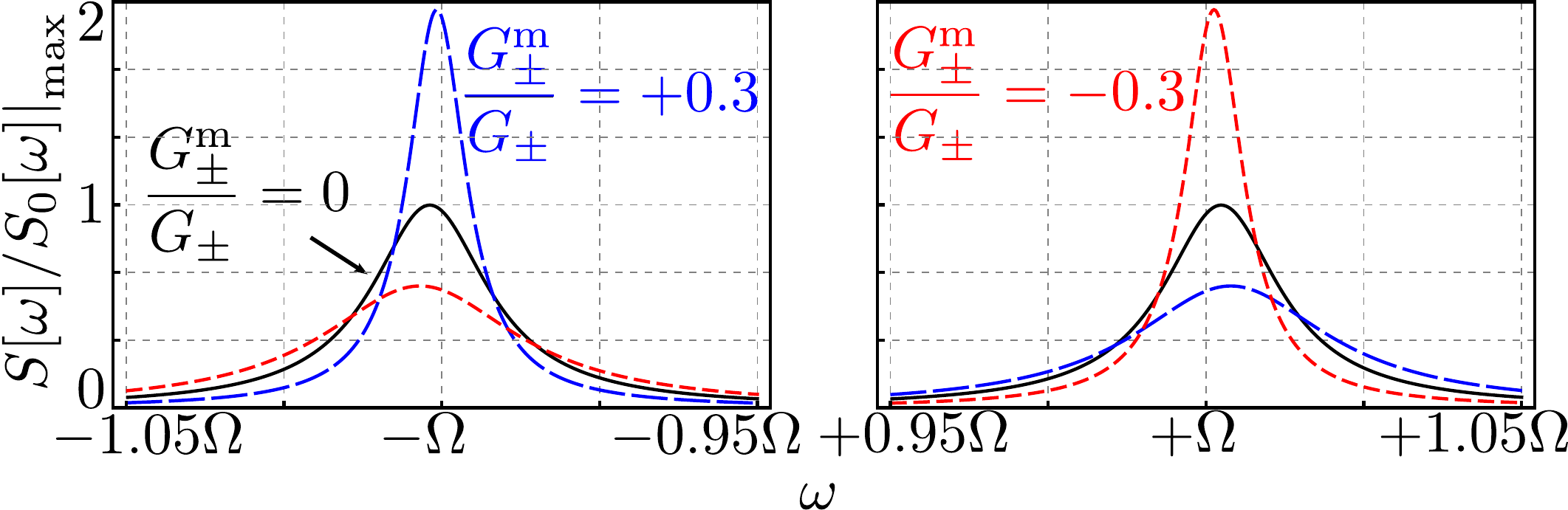} 
\end{center}
\caption{  (Color online) Cavity output spectra, as defined in Eq.~(\ref{eq:outputspectrumdef}), centred around detunings from the cavity resonance frequency of (left panel) $\omega = -\Omega$ and (right panel) $\omega = +\Omega$. The spectra are shown for the case (solid black curve) without imperfections in the effective couplings $(G^{\rm m}_{\pm} = 0)$, and for the cases $G^{\rm m}_{\pm}/G_{\pm} = 0.3$ (blue curve with long dashes) and $G^{\rm m}_{\pm}/G_{\pm} = -0.3$ (red curve with short dashes). Imperfections in the effective couplings lead to asymmetry in the observed output spectra, see Eq.~(\ref{eq:spectralpeaks}). In the absence of these imperfections, the steady-state mechanical entanglement can be bounded based on a measurement of the output spectrum. The spectra are shown for a drive asymmetry $G_+/G_- = 0.9$ and a cooperativity $C_- = 1200$, while other parameters are $\bar{n} = 0$, $\kappa = 2\pi \times 1.592 \times 10^5 \, \rm{Hz} = 10^6 \, \rm{s^{-1}}$, $\gamma/\kappa = 4\times 10^{-5}$ and $\Omega/\kappa = 0.1$. } 
\label{fig:spectra}
\end{figure*}

\subsection{Output spectrum}
From an experimental point of view, reconstructing the entire covariance matrix would be extremely demanding. Even performing direct measurements of both of the collective quadatures required for testing the Duan criterion of Eq.~(\ref{eq:Duan}) would be difficult. However, one could perform a back-action-evading measurement of the collective quadrature $\hat{X}_+$, and take this as some evidence for the existence of two-mode squeezing in the steady-state. Alternatively, we can seek a signature of the mechanical entanglement in the cavity output spectrum. As usual, the output spectrum is calculated by first solving the Heisenberg-Langevin equations in the frequency domain. Taking the Fourier transform of Eq.~(\ref{eq:HLeqns}) we find  
\begin{equation}
\vec{X} [\omega ] = - (\mathbf{A}_0 + i\omega \mathbf{I}_6 )^{-1} \cdot \mathbf{B}_0 \cdot \vec{X} [ \omega ] , \label{eq:HLfrequency}
\end{equation}
where $\vec{X}[\omega ] = ( \hat{a}[\omega ] , \hat{a}^\dagger [\omega ] , \hat{b}[ \omega ], \hat{b}^\dagger [\omega ], \hat{c}[\omega ], \hat{c}^\dagger [\omega ] )^T $ and the matrices are given in App.~\ref{eq:modeops}. The output spectrum is calculated in the standard manner \cite{walls} as
\begin{equation}
S[\omega ] = \int dt \, e^{i\omega t} \langle \delta \hat{c}^\dagger_{\rm out}(t) \delta \hat{c}_{\rm out} (0) \rangle, \label{eq:outputspectrumdef}
\end{equation}
where the output cavity field is given by $\delta \hat{c}_{\rm out} = \hat{c}_{\rm out} - \langle \hat{c}_{\rm out} \rangle$ and $\hat{c}_{\rm out} = \hat{c}_{\rm in} + \sqrt{\kappa} \hat{c}_{\rm in}$. 

We first calculate the spectrum assuming that there are no imperfections in the effective couplings $(G^{\rm m}_{\pm} = 0)$ and ignoring time-dependent Hamiltonian contributions; that is, with the effective Hamiltonian (\ref{eq:HamGen1}). Then the cavity output spectrum, in the limit $\gamma/\Omega \rightarrow 0$, is given by
\begin{equation}
S[\omega ] = \kappa \frac{ 32 \left[ G^2_- \bar{n} + G^2_+ ( \bar{n} + 1)  \right] \gamma \left[ \gamma^2 + 4(\omega^2 + \Omega^2) \right] }{ \left| N(\omega ) \right|^2 }, 
\end{equation}
where $N(\omega ) = \left[ 8\mathcal{G}^2 + (\gamma - 2i\omega ) (\kappa - 2i\omega ) \right] (\gamma - 2i\omega ) + 4\Omega^2 (\kappa - 2i\omega )$. In the case $\Omega = 0$ this reduces to the result for a single mechanical oscillator \cite{kronwald}. As shown in Fig.~\ref{fig:spectra}, the output spectrum exhibits peaks at detunings around $\pm \Omega$ from the cavity resonance frequency. This corresponds to the drive photons being scattered towards the cavity resonance, with an energy $\omega_a$ or $\omega_b$ being provided by or extracted from the mechanical oscillators. As the optomechanical damping rate $\Gamma$ is increased, the widths of the spectral peaks increase and they are shifted to larger detunings (for $G_-/G_+ > 0$), as expected.  

If we now allow for the possibility of imperfections in the effective couplings, but still ignore time-dependent contributions, the cavity output spectrum is again readily obtained. The corresponding spectra are also shown in Fig.~\ref{fig:spectra}. The general expression is complicated, but at detunings of $\pm \Omega$ we find
\begin{equation}
S[ \pm \Omega ] = \gamma \kappa \frac{ (G_- \pm G^{\rm m}_-)^2 \bar{n} + (G_+ \pm G^{\rm m}_+ )^2 (1+\bar{n}) }{ \left[ G^2_- - (G^{\rm m}_-)^2 - G^2_+ + (G^{\rm m}_+)^2 \right]^2 } . \label{eq:spectralpeaks}
\end{equation}
Clearly, the asymmetry in the spectral peaks is determined by the imperfections,  $G^{\rm m}_{\pm}$, in the effective optomechanical couplings.  

\subsection{Bogoliubov modes}
Knowledge of the cavity output spectrum can be used to provide us with information about the occupations of the mechanical Bogoliubov modes. In particular, neglecting imperfections in the effective couplings and in the limit $\gamma/\Omega \rightarrow 0$, we can show that the occupations of the Bogoliubov modes are related to the integral of each peak in the output spectrum by
\begin{eqnarray}
\int^0_{-\infty} S[\omega ] d\omega & = & \int^{+\infty}_0 S[\omega ] d\omega \nonumber \\
& = & 8\pi \kappa \frac{ \mathcal{G}^2 }{ 4\mathcal{G}^2 + \kappa (\kappa + \gamma ) } \langle \hat{\beta}^\dagger_i \hat{\beta}_i \rangle , \label{eq:integratedspectrum}
\end{eqnarray}
for $i=1$ or $2$ (in this, and subsequent, expressions). Therefore, from the output spectrum and knowledge of the system parameters one can determine the occupations of the Bogoliubov modes. The same information can be obtained from the heights of the spectral peaks, since (in the same limit) we also have
\begin{eqnarray}
S[ \pm \Omega ] & = & \frac{ \gamma \kappa + 4(G^2_- - G^2_+) }{ G^2_- - G^2_+ } \langle \hat{\beta}^\dagger_i \hat{\beta}_i \rangle . \label{eq:peakheight}
\end{eqnarray}
Note that Eqs.~(\ref{eq:integratedspectrum}) and (\ref{eq:peakheight}) only hold when the imperfections in the effective couplings are less than $\sim 1\%$. A similar result to that of Eq.~(\ref{eq:integratedspectrum}) was obtained for a single mechanical oscillator \cite{kronwald}, though in that case the integration is over all frequencies. 

\subsection{Entanglement criterion }
Now from Eqs.~(\ref{eq:integratedspectrum}) and (\ref{eq:peakheight}) it is clear that we can estimate the occupations of the Bogoliubov modes using the cavity output spectrum. Recall that the simplest means of verifying the presence of mechanical entanglement is via the Duan criterion, Eq.~(\ref{eq:Duan}). The task then is to bound the Duan quantity using our knowledge of the occupations of the Bogoliubov modes.

Repeated application of the generalized Cauchy-Schwarz inequality \cite{glauber,kronwald}, allows one to show that $ | \langle \hat{\beta}^2_i \rangle | \leq \langle \hat{\beta}^\dagger_i \hat{\beta}_i \rangle + 1/2 $. With the additional assumption that $\langle \hat{\beta}^\dagger_1 \hat{\beta}_1 \rangle = \langle \hat{\beta}^\dagger_2 \hat{\beta}_2 \rangle $, known to be true in the absence of imperfections in the couplings and in the limit $\gamma/\Omega \rightarrow 0$, we can bound the Duan quantity. Explicitly, we find that 
\begin{equation}
\langle \hat{X}^2_+ \rangle + \langle \hat{P}^2_- \rangle \leq 8 e^{-2r} \left( \langle \hat{\beta}^\dagger_i \hat{\beta}_i \rangle + 1/2 \right) .
\end{equation}
The parameter $r$ is known from the drive asymmetry, c.f. Eq.~(\ref{eq:r}). The Duan quantity and its bound converge in the highly-entangled (large-$r$) regime, and therefore we expect it to reliably indicate the existence of an entangled mechanical steady-state.   

\section{Two cavity modes, one mechanical oscillator}

\begin{figure}[ht]
\begin{center}
\includegraphics[scale=0.38]{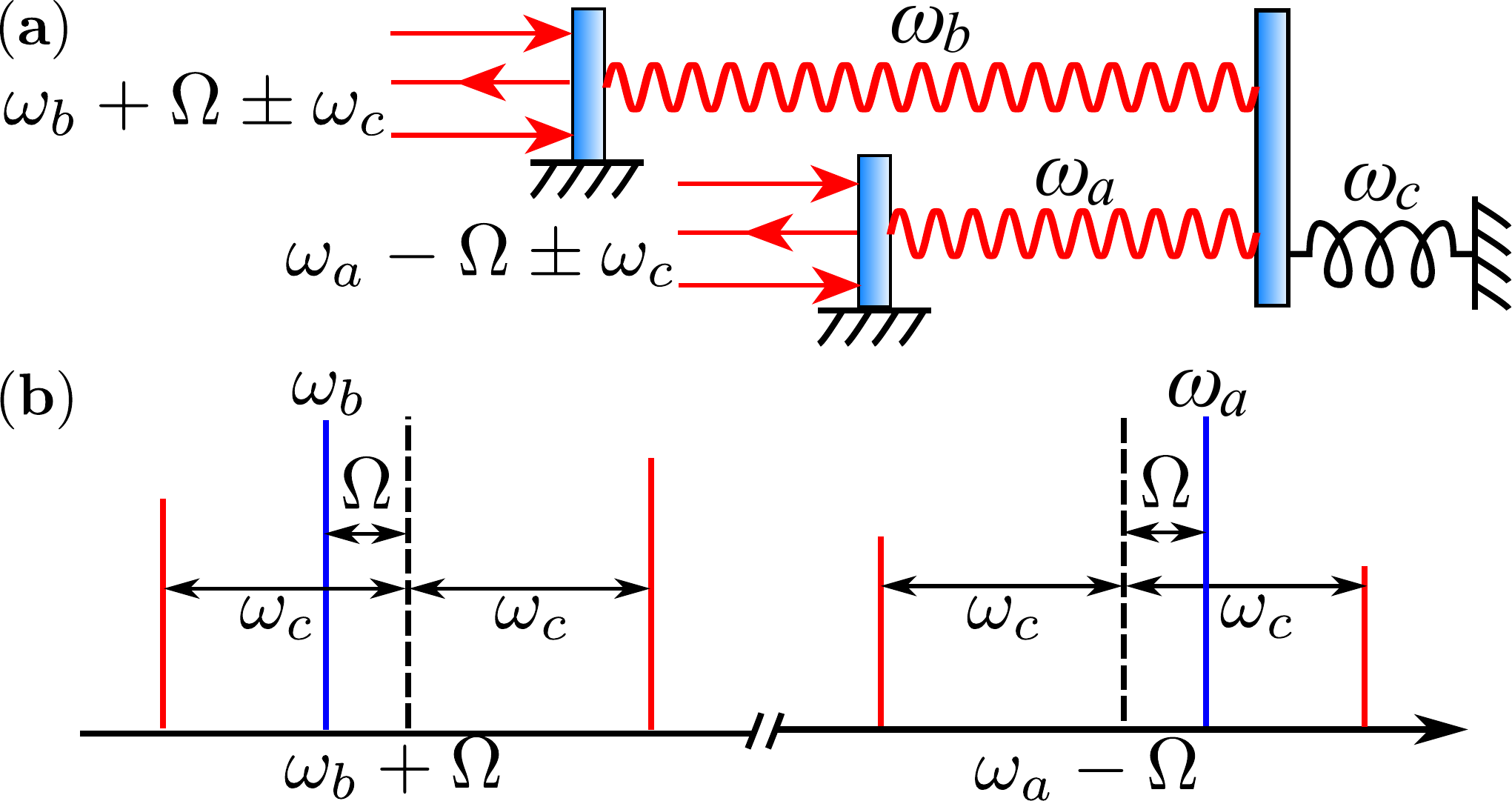}
\end{center}
\caption{  (Color online) (a) Representation of a three-mode optomechanical system composed of two cavity modes (with resonance frequencies $\omega_a$ and $\omega_b$), each independently coupled by radiation pressure to a single mechanical oscillator (with resonance frequency $\omega_c$). (b) The driving conditions, in terms of frequencies, that lead to the linearised effective Hamiltonian (\ref{eq:HamGen1}) with the role of the cavity and mechanical modes interchanged. The cavity resonance frequencies are indicated by the blue lines, while the driving tones are indicated by red lines. The driving tones are placed symmetrically (at detunings $\pm \omega_c$) about detunings of $\Omega$ from the cavity resonance frequencies. } 
\label{fig:twocavities}
\end{figure}

\label{sec:cavity}
Thus far we have considered a three-mode optomechanical system composed of two mechanical oscillators coupled to one common cavity mode. It is of considerable interest, particularly from the perspective of quantum information processing, to consider the opposite scenario in which there are two cavity modes coupled to a single mechanical oscillator \cite{barzanjeh,wang1,cleland,lehnert2}. In particular, the ubiquity of optomechanical couplings raises the possibility of entangling cavity modes of vastly different frequencies (e.g. microwave and optical modes). This three-mode optomechanical system is depicted schematically in Fig.~\ref{fig:twocavities}(a). 

Now the Hamiltonian of the system is, c.f. Eq.~(\ref{eq:HamOne}),
\begin{eqnarray}
\hat{\mathcal{H}} & = & \omega_a \hat{a}^\dagger \hat{a} + \omega_b \hat{b}^\dagger \hat{b} + \omega_c \hat{c}^\dagger \hat{c} + g_a (\hat{c} + \hat{c}^\dagger ) \hat{a}^\dagger \hat{a} \nonumber \\
& & + g_b (\hat{c} + \hat{c}^\dagger ) \hat{b}^\dagger \hat{b} + \hat{H}_{\rm drive} + \hat{H}_{\rm diss} , \label{eq:twocavity}
\end{eqnarray}
where $\hat{a}$ and $\hat{b}$ describe two cavity modes, and $\hat{c}$ describes the mechanical oscillator. We shall assume a driving Hamiltonian of the form
\begin{eqnarray}
\hat{H}_{\rm drive} & = & \left( \mathcal{E}_{a+} e^{+i\omega_c t} + \mathcal{E}_{a-} e^{-i\omega_c t} \right) e^{+i ( \omega_a - \Omega ) t} \hat{a} \nonumber \\
& & + \left( \mathcal{E}_{b+} e^{+i\omega_c t} + \mathcal{E}_{b-} e^{-i\omega_c t} \right)  e^{+i( \omega_b + \Omega ) t} \hat{b} + {\rm h.c.} \nonumber \\
& & 
\end{eqnarray}
That is, driving at $\omega_a - \Omega \pm \omega_c$ and $\omega_b + \Omega \pm \omega_c$ for cavity modes $\hat{a}$ and $\hat{b}$, respectively, as depicted in Fig.~\ref{fig:twocavities}(b). In an interaction picture defined with respect to the Hamiltonian (\ref{eq:framedef}), the Hamiltonian (\ref{eq:twocavity}) takes the form of the Hamiltonian (\ref{eq:HamGen1}) if we set $g_a \bar{a}_+  = g_b \bar{b}_+ \equiv G_+$ and $g_a \bar{a}_- = g_b \bar{b}_- \equiv G_-$, with $\bar{a}_{\pm}$ and $\bar{b}_{\pm}$ denoting the steady-state amplitudes at the driven cavity sidebands. Thus we can realize the same physics with the cavity modes that we described previously for the mechanical modes, including the possibility of generating highly-pure, highly-entangled electromagnetic modes. As before, this derivation relies on the assumptions that we are operating in the resolved-sideband regime, that the driving strengths at the driven sidebands are large, and that the effects of counter-rotating terms are negligible. The deviations arising from abandoning any of these assumptions, may of course, be calculated. 

\section{Conclusions}
We have provided a detailed proposal for configuring a three-mode optomechanical system such that the steady-state includes a highly-pure, highly-entangled two-mode squeezed state. The generation of both mechanical and electromagnetic two-mode squeezed states has been described. The symmetry of this steady-state makes it an attractive platform for the implementation of continuous-variable teleportation protocols. The proposal is efficient in the sense that it requires only one driven auxiliary mode to be configured as the engineered reservoir. The problem of unequal single-photon optomechanical couplings has been overcome by using a four-tone driving scheme, and potential instabilities arising from counter-rotating terms have been accounted for. A simple experimental signature for the presence of mechanical entanglement, in terms of the cavity output spectrum, has been provided. The proposal described is implementable with existing technology. 

\section{Acknowledgements}
This work was supported by NSERC, an ECR Grant from UNSW Canberra, and the DARPA ORCHID program under a grant from the AFOSR. We thank Junho Suh, Chan U Lei, Ian Petersen and Seiji Armstrong for useful discussions. 

\appendix

\section{Derivation of Hamiltonian }
\label{sec:derivations}
We consider the three-mode optomechanical system depicted in Fig.~\ref{fig:optodualmechanics}(a) and \ref{fig:optomechanics4tone}(a), and start from the Hamiltonian (\ref{eq:HamOne}) with the four-tone drive (\ref{eq:fourdrives}); the scenario with the two-tone drive (\ref{eq:twotone}) shall be treated as a special case. Moving into a rotating frame with respect to $\hat{H}_0 = \omega_1 \hat{a}^\dagger \hat{a} + \omega_2 \hat{b}^\dagger \hat{b} + \omega_c \hat{c}^\dagger \hat{c}$, recalling Eqs.~(\ref{eq:shorthand1}) and (\ref{eq:shorthand2}), we obtain 
\begin{eqnarray}
\hat{\mathcal{H}} & = & \Omega ( \hat{a}^\dagger \hat{a} - \hat{b}^\dagger \hat{b} ) + g_a ( \hat{a} e^{-i\omega_1 t} + \hat{a}^\dagger e^{+i\omega_1 t} ) \hat{c}^\dagger \hat{c} \nonumber \\ 
& & + g_b ( \hat{b} e^{-i\omega_2 t} + \hat{b}^\dagger e^{+i\omega_2 t} ) \hat{c}^\dagger \hat{c} + \hat{H}_{\rm drive} + \hat{H}_{\rm diss} . \nonumber \\ 
& & \label{eq:HamRotating}
\end{eqnarray}
The effective oscillation frequency $\Omega $ is chosen such that Eqs.~(\ref{eq:OmegaGamma}) and (\ref{eq:unwantedSB}) are satisfied. The Heisenberg equations, neglecting noise terms, corresponding to Eq.~(\ref{eq:HamRotating}) are
\begin{subequations}
\begin{eqnarray}
\dot{\hat{a}} & = & -i\Omega \hat{a} - ig_a e^{+ i \omega_1 t} \hat{c}^\dagger \hat{c} - \frac{\gamma_a }{2} \hat{a} \label{eq:O1} , \\
\dot{\hat{b}} & = & i\Omega \hat{b} - ig_b e^{+ i \omega_2 t} \hat{c}^\dagger \hat{c} - \frac{\gamma_b}{2} \hat{b} , \\
\dot{\hat{c}} & = & -ig_a \left( \hat{a} e^{-i \omega_1 t} + \hat{a}^\dagger e^{+ i \omega_1 t } \right) \hat{c}  - i\mathcal{E}_{1+} e^{-i \omega_1 t} \nonumber \\ 
& & - i\mathcal{E}_{1-} e^{+i \omega_1 t} - ig_b \left( \hat{b} e^{-i \omega_2 t} + \hat{b}^\dagger e^{+i \omega_2 t} \right) \hat{c} \nonumber \\
& & - i\mathcal{E}_{2+} e^{-i \omega_2 t}  - i\mathcal{E}_{2-} e^{+i \omega_2 t} - \frac{\kappa}{2} \hat{c} . \label{eq:O3}
\end{eqnarray}
\end{subequations}
Assuming resolved-sideband operation ($\omega_{a(b)} \gg \kappa$), we take the ansatz \cite{woolley:nanomechanicalsqueezing} 
\begin{eqnarray}
\hat{c}(t) & = & \hat{c}_0 (t) + \hat{c}_{1+}(t) e^{-i\omega_1 t} + \hat{c}_{1-}(t) e^{+i\omega_1 t} \nonumber \\
& & + \hat{c}_{2+} (t) e^{-i\omega_2 t} + \hat{c}_{2-}(t) e^{+i\omega_2 t} . \label{eq:ansatz}
\end{eqnarray}
Then substituting Eq.~(\ref{eq:ansatz}) into the system (\ref{eq:O1})-(\ref{eq:O3}) and separating out the Fourier coefficients of the cavity field we obtain the system:
\begin{widetext}
\begin{subequations}
\begin{eqnarray}
\dot{\hat{a}} & = & -i\Omega \hat{a} -ig_a \left( \hat{c}^\dagger_{1-} \hat{c}_0 + \hat{c}_{1+} \hat{c}^\dagger_0 \right) -\frac{\gamma_a}{2} \hat{a} - ig_a e^{+2i\delta t} \left( \hat{c}^\dagger_{2-} \hat{c}_0 + \hat{c}_{2+} \hat{c}^\dagger_0 \right) \nonumber \\
& & -ig_a e^{+2i\omega_1 t} \left( \hat{c}^\dagger_{1+} \hat{c}_0 + \hat{c}_{1-} \hat{c}^\dagger_0 \right) - ig_a e^{+2i\omega_m t} \left( \hat{c}^\dagger_{2+} \hat{c}_0 + \hat{c}_{2-} \hat{c}^\dagger_0 \right) , \label{eq:compa} \\
\dot{\hat{b}} & = &i\Omega \hat{b} -ig_b   \left( \hat{c}^\dagger_{2-} \hat{c}_0 + \hat{c}_{2+} \hat{c}^\dagger_0 \right)  -\frac{\gamma_b}{2} \hat{b} -ig_b e^{-2i\delta t} \left( \hat{c}^\dagger_{1-} \hat{c}_0 + \hat{c}_{1+} \hat{c}^\dagger_0 \right) \nonumber \\
& & -ig_b e^{+2i \omega_m t} \left( \hat{c}^\dagger_{1+} \hat{c}_0 + \hat{c}_{1-} \hat{c}^\dagger_0 \right) - ig_b e^{+2i\omega_2 t} \left( \hat{c}^\dagger_{2+} \hat{c}_0 + \hat{c}_{2-} \hat{c}^\dagger_0 \right) , \label{eq:compb} \\
\dot{\hat{c}}_0 & = & -ig_a \hat{a} \hat{c}_{1-} - ig_a \hat{a}^\dagger \hat{c}_{1+} - ig_b \hat{b} \hat{c}_{2-} - ig_b \hat{b}^\dagger \hat{c}_{2+} - \frac{\kappa}{2} \hat{c}_0 \nonumber \\
& & -ig_a \left( \hat{a} \hat{c}_{2-} e^{-2i\delta t} + \hat{a}^\dagger \hat{c}_{2+} e^{+2i\delta t} \right) - ig_b \left( \hat{b} \hat{c}_{1-} e^{+2i\delta t} + \hat{b}^\dagger \hat{c}_{1+} e^{-2i\delta t}  \right) \nonumber \\
& & -ig_a \left( \hat{a} \hat{c}_{1+} e^{-2i\omega_1 t} + \hat{a} \hat{c}_{2+} e^{-2i \omega_m t} + \hat{a}^\dagger \hat{c}_{1-} e^{+2i\omega_1 t} + \hat{a}^\dagger \hat{c}_{2-} e^{+2 i \omega_m t} \right) \nonumber \\
& & - ig_b \left( \hat{b} \hat{c}_{1+} e^{-2i \omega_m t} + \hat{b} \hat{c}_{2+} e^{-2i\omega_2 t} + \hat{b}^\dagger \hat{c}_{1-} e^{+2i \omega_m t} + \hat{b}^\dagger \hat{c}_{2-} e^{+2i\omega_2 t} \right)  , \label{eq:compc} \\
\dot{\hat{c}}_{1-} & = & (-i\omega_1 - \kappa /2) \hat{c}_{1-} - ig_a \hat{a}^\dagger \hat{c}_0 - i\mathcal{E}_{1-} \label{eq:F4} , \\
\dot{\hat{c}}_{1+} & = & (i\omega_1 - \kappa /2) \hat{c}_{1-} - ig_a \hat{a} \hat{c}_0 - i\mathcal{E}_{1+}  , \\
\dot{\hat{c}}_{2-} & = & (-i\omega_2 - \kappa /2) \hat{c}_{1-} - ig_b \hat{b}^\dagger \hat{c}_0 - i\mathcal{E}_{2-}  , \\
\dot{\hat{c}}_{2+} & = & (i\omega_2 - \kappa /2) \hat{c}_{1-} - ig_b \hat{b} \hat{c}_0 - i\mathcal{E}_{2+} \label{eq:F7} .
\end{eqnarray}
\end{subequations}
\end{widetext}
Note that in writing out Eqs.~(\ref{eq:compa})-(\ref{eq:F7}) we have retained fast-rotating terms only in Eqs.~(\ref{eq:compa})-(\ref{eq:compc}). Solving Eqs.~(\ref{eq:F4})-(\ref{eq:F7}) for the steady-state amplitudes of the field at the driven sidebands, assuming that the single-photon optomechanical couplings are relatively small, we obtain the results of Eq.~(\ref{eq:SBamplitudes}).

Replacing the operator Fourier components at the driven sidebands by their classical steady-state values, we can write down an effective (quadratic) Hamiltonian for the system dynamics. If we neglect fast-rotating terms in Eqs.~(\ref{eq:compa})-(\ref{eq:compc}), this Hamiltonian will be time-independent; the time-dependent contributions to the effective Hamiltonian are given in App.~\ref{sec:HamCR}.

With four-tone driving, the effective Hamiltonian is (replacing $\hat{c}_0 \rightarrow \hat{c}$),
\begin{eqnarray}
\hat{\mathcal{H}} & = &  \Omega \left( \hat{a}^\dagger \hat{a} - \hat{b}^\dagger \hat{b} \right) + g_a \left[ ( \bar{c}_{1-} \hat{a} + \bar{c}_{1+} \hat{a}^\dagger ) \hat{c}^\dagger + {\rm h.c.} \right] \nonumber \\
& & + g_b \left[ ( \bar{c}_{2-} \hat{b} + \bar{c}_{2+} \hat{b}^\dagger ) \hat{c}^\dagger + {\rm h.c. } \right] + \hat{H}_{\rm diss} . \label{eq:effHam4}
\end{eqnarray}
Assuming that the drives are matched according to Eq.~(\ref{eq:matchingcondition}) and that the steady-state amplitudes in the driven sidebands are real, we get the Hamiltonian (\ref{eq:HamGen1}) with the effective couplings of Eq.~(\ref{eq:GPlusMinus}). If there is an imperfection (drive mismatch), then we have the additional Hamiltonian contributions given by Eq.~(\ref{eq:HamBetter}) with the coupling imperfections given by Eq.~(\ref{eq:coupleimperfect2}). 

With two-tone driving we have $\omega_1 = \omega_2 = \omega_m$, and the appropriate effective Hamiltonian is now (\ref{eq:effHam4}) with $\bar{c}_{k\pm}$ replaced by $\bar{c}_{\pm}$, the driving strengths at the frequencies $\omega_c \pm \omega_m$. If the single-photon optomechanical couplings are equal, then the effective Hamiltonian is given by Eq.~(\ref{eq:HamGen1}) where the effective couplings are given by Eq.~(\ref{eq:optomechTwo}). Imperfections (unequal single-photon optomechanical couplings) lead to additional contributions of the form of Eq.~(\ref{eq:HamBetter}), where the effective coupling imperfections are given by Eq.~(\ref{eq:optomechTwo}). 

\section{Two-mode Gaussian states}
\label{sec:TMGS}
\subsection{Entanglement }
\label{sec:entanglementandpurity}
The entanglement of a two-mode Gaussian state may be quantified from its symmetrically-ordered covariance matrix, $\mathbf{V}$, via the logarithmic negativity \cite{vidal,adesso,plenio}. Writing the covariance matrix as in Eq.~(\ref{eq:blockform}), the logarithmic negativity is given by   
\begin{equation}
E_{\mathcal{N}} = \max \left\{ 0,- \ln 2\eta \right\}, 
\end{equation}
where $\eta = 2^{-1/2} \{  \Sigma (\mathbf{V}) - \left[ \Sigma (\mathbf{V})^2 - 4 {\rm det} \ \mathbf{V} \right]^{1/2} \}^{1/2} $ and $\Sigma ( \mathbf{V} ) = {\rm det} \ \mathbf{V}_b + {\rm det} \ \mathbf{V}_a -2 \, {\rm det} \mathbf{V}_{ab} $. 

\subsection{Thermal two-mode squeezed state}
\label{sec:TTMSS}
An alternative method for characterising the mechanical two-mode state is obtained by noting that the covariance matrix of our steady-state takes the form of a \emph{thermal} two-mode squeezed state \cite{marian}, as defined in Eq.~(\ref{eq:thermalTMSSmain}). This state is described by three parameters: a two-mode squeezing parameter $\xi$, and the two initial thermal occupations, $\bar{n}^{a}_{\rm th}$ and $\bar{n}^{b}_{\rm th}$. The covariance matrix for the thermal two-mode squeezed state, in the block form of Eq.~(\ref{eq:blockform}), is 
\begin{equation}
\mathbf{V}_a = c_a  \mathbf{I}_{2}, \, \, \mathbf{V}_b = c_b \mathbf{I}_{2}, \, \, \mathbf{V}_{ab} = -c_{ab} \sigma_z, 
\end{equation}
where the coefficients are given by
\begin{subequations}
\begin{eqnarray}
c_{a (b)} & = & ( \bar{n}^{a (b)}_{\rm th} + 1/2 ) \cosh^2 \xi + ( \bar{n}^{b (a)}_{\rm th} + 1/2 ) \sinh^2 \xi , \nonumber \\
& & \\
c_{ab} & = & \left( \bar{n}^{a}_{\rm th} + \bar{n}^b_{\rm th} + 1 \right) \sinh \xi \cosh \xi . 
\end{eqnarray}
\end{subequations}

\section{Heisenberg-Langevin equations}

\subsection{Collective quadratures}
\label{sec:HLcoll}
In the adiabatic limit and in terms of the collective quadratures of Eqs.~(\ref{eq:collectivequad1}) and (\ref{eq:collectivequad2}), the dynamics of the two-mode mechanical system are described by Eq.~(\ref{eq:adiabaticlangevin}), with the matrices
\begin{subequations}
\begin{eqnarray}
\mathbf{A}_0 & = & \left[ \begin{array}{c|c} \mathbf{A}_+ & \mathbf{A}_{+-} \\ \hline \mathbf{A}_{+-} & \mathbf{A}_- \end{array} \right] , \\
\mathbf{B}_{1} & = & \left[ \begin{array}{c|c} \mathbf{B}_{1+} & \mathbf{B}_{1-} \\ \hline \mathbf{B}_{1+} & \mathbf{B}_{1-} \end{array} \right] , \\
\mathbf{B}_2 & = & \sqrt{\Gamma} \left[ \begin{array}{c} \mathbf{I}_2 \\ \hline \mathbf{0}_{22} \end{array} \right] , 
\end{eqnarray}
\end{subequations}
including the components 
\begin{subequations}
\begin{eqnarray}
\mathbf{A}_+ & = & - (\gamma/2 + \Gamma ) \mathbf{I}_2, \\
\mathbf{A}_- & = & -(\gamma/2) \mathbf{I}_2, \\
\mathbf{B}_{1\pm} & = & \sqrt{ \gamma (1 \pm l)/2 } \mathbf{I}_2, \\
\mathbf{A}_{+-} & = & \left[ \begin{array}{cc} -l\gamma/2 & \Omega \\ -\Omega & -l\gamma/2 \end{array} \right] ,
\end{eqnarray}
\end{subequations}
where $l = (\gamma_a - \gamma_b)/(2\gamma)$ and $\gamma = (\gamma_a + \gamma_b)/2$. 

\subsection{Individual quadratures}
\label{sec:HLind}
The dynamics of the linearised, three-mode optomechanical system, with Hamiltonian (\ref{eq:HamGen1}), are described by Eq.~(\ref{eq:HLeqns}). The system matrix is
\begin{equation}
\mathbf{A}_0 = \left[ \begin{array}{c|c|c} \mathbf{A}_a & \mathbf{0}_{22} & \mathbf{C}_a \\ \hline \mathbf{0}_{22} & \mathbf{A}_b & \mathbf{C}_b \\ \hline \mathbf{C}_a & \mathbf{C}_b & \mathbf{A}_c \end{array} \right] , \label{eq:A0Block}
\end{equation}
where $\mathbf{0}_{22}$ is the $2\times 2$ zero matrix, and 
\begin{subequations}
\begin{eqnarray}
\mathbf{A}_a & = & \left[ \begin{array}{cc} -\gamma_a/2 & \Omega \\ -\Omega & -\gamma_a/2 \end{array} \right] , \\
\mathbf{A}_b & = & \left[ \begin{array}{cc} -\gamma_b/2 & -\Omega \\ \Omega & -\gamma_b/2 \end{array} \right] , \\
\mathbf{A}_c & = & -(\kappa/2) \mathbf{I}_2, \label{eq:Ac} \\ 
\mathbf{C}_a & = & \left[ \begin{array}{cc} 0 & G_- - G_+ - G^{\rm m}_{\rm s} \\ - G_- - G_+ + G^{\rm m}_{\rm d} & 0 \end{array} \right] , \nonumber \\
& & \\
\mathbf{C}_b & = & \left[ \begin{array}{cc} 0 & G_- - G_+ + G^{\rm m}_{\rm s} \\ - G_- - G_+ - G^{\rm m}_{\rm d} & 0 \end{array} \right] , \nonumber \\
& & 
\end{eqnarray}
\end{subequations}
with the short-hand notation, $G^{\rm m}_{\rm s} = G^{\rm m}_- + G^{\rm m}_+$ and $G^{\rm m}_{\rm d} = G^{\rm m}_- - G^{\rm m}_+$. The noise matrix is given by 
\begin{eqnarray}
\mathbf{B}_0 & = & {\rm Diag} \left( \sqrt{ \gamma_a (\bar{n}_a + 1/2) } \mathbf{I}_2, \right. \nonumber \\
& & \left. \sqrt{ \gamma_b (\bar{n}_b + 1/2) } \mathbf{I}_2 ,  \sqrt{\kappa/2} \mathbf{I}_2 \right) . 
\end{eqnarray}

\subsection{Individual mode operators}
\label{eq:modeops}
For the purpose of evaluating the cavity output spectrum, it is more convenient to work in terms of annihilation and creation operators, as in Eq.~(\ref{eq:HLfrequency}). The corresponding matrices are
\begin{equation}
\mathbf{B}_0 = {\rm Diag} \left( \sqrt{\gamma_a} \mathbf{I}_2, \sqrt{\gamma_b} \mathbf{I}_2, \sqrt{\kappa} \mathbf{I}_2 \right) , 
\end{equation}
while $\mathbf{A}_0$ is given by the block matrix form of Eq.~(\ref{eq:A0Block}), now with 
\begin{subequations}
\begin{eqnarray}
\mathbf{A}_a & = & \left[ \begin{array}{cc} -i\Omega - \gamma_a/2 & 0 \\ 0 & i\Omega - \gamma_a/2 \end{array} \right] , \\
\mathbf{A}_b & = & \left[ \begin{array}{cc} i\Omega - \gamma_b/2 & 0 \\ 0 & -i\Omega - \gamma_b/2 \end{array} \right] , 
\end{eqnarray}
\begin{eqnarray}
\mathbf{C}_a & = & i \left[ \begin{array}{cc} -G_- + G^{\rm m}_- & -G_+ - G^{\rm m}_+ \\  G_+ + G^{\rm m}_+ & G_- - G^{\rm m}_- \end{array} \right] , \\
\mathbf{C}_b & = & i \left[ \begin{array}{cc} -G_- - G^{\rm m}_- & -G_+ + G^{\rm m}_+ \\ G_+ - G^{\rm m}_+ & G_- + G^{\rm m}_- \end{array} \right] ,
\end{eqnarray}
\end{subequations}
while $\mathbf{A}_c$ is still given by Eq.~(\ref{eq:Ac}). 

\section{Counter-rotating contributions}
\label{sec:CR}

\subsection{Hamiltonians}
\label{sec:HamCR}

In deriving the time-independent Hamiltonian (\ref{eq:HamGen1}) we discarded fast-rotating terms; here we include them. The full time-dependent Hamiltonian is 
\begin{equation}
\hat{\mathcal{H}}(t) = \hat{\mathcal{H}} + \hat{\mathcal{H}}_{CR} , \label{eq:Htot}
\end{equation}
where $\hat{\mathcal{H}}$ is the time-independent effective Hamiltonian (\ref{eq:HamGen1}) and $\hat{\mathcal{H}}_{CR}$ is the time-dependent (``counter-rotating'') contribution. 

\subsubsection{Four-tone Driving}
With four driving tones (\ref{eq:fourdrives}) the time-dependent part of the Hamiltonian (\ref{eq:Htot}) is
\begin{eqnarray}
\hat{\mathcal{H}}_{CR} & = & g_a \left\{ \hat{a} \left[ \bar{c}_{2-} e^{-2i\delta t} + \bar{c}_{2+} e^{-2i\omega_m t}  \right. \right. \nonumber \\ 
& & \left. + \bar{c}_{1+} e^{-2i\omega_1 t} \right] + \hat{a}^\dagger \left[ \bar{c}_{2+} e^{+2i\delta t} + \bar{c}_{1-} e^{+2i\omega_1 t} \right. \nonumber \\ 
& & \left. \left. + \bar{c}_{2-} e^{2i \omega_m t} \right] \right\} \hat{c}^\dagger \nonumber \\
& & + g_b \left\{ \hat{b} \left[ \bar{c}_{1-} e^{+2i\delta t} + \bar{c}_{1+} e^{-2i \omega_m t} \right. \right. \nonumber \\
& & \left. + \bar{c}_{2+} e^{-2i\omega_2 t} \right] + \hat{b}^\dagger \left[ \bar{c}_{1+} e^{-2i\delta t} + \bar{c}_{2-} e^{+2i\omega_2 t} \right. \nonumber \\
& & \left. \left. + \bar{c}_{1-} e^{+2i \omega_m t} \right] \right\} \hat{c}^\dagger  + {\rm h.c.} \label{eq:CRfour}
\end{eqnarray}
There are terms at four distinct oscillation frequency magnitudes: the $\pm 2 \delta $ terms are associated with the two drives being on the same side of the cavity resonance frequency, while the terms oscillating at $\pm 2( \omega_a- \Omega ), \pm 2\omega_m, \pm2(\omega_b + \Omega ) $ are associated with two drives on opposing sides of the cavity resonance frequency. In terms of Bogoliubov modes, the Hamiltonian (\ref{eq:CRfour}) may be written out as
\begin{widetext}
\begin{eqnarray}
\hat{\mathcal{H}}_{CR} & = & \mathcal{G} e^{-2i\delta t} \left[ \hat{\beta}_1 \hat{c}^\dagger \left( 1/d + (d-1/d)\cosh^2 r \right) - \hat{\beta}^\dagger_2 \hat{c}^\dagger (d-1/d) \cosh r \sinh r\right] \nonumber \\
& & + \mathcal{G} e^{+2i\delta t} \left[ \hat{\beta}_2 \hat{c}^\dagger \left( 1/d - (d-1/d) \sinh^2 r \right) + \hat{\beta}^\dagger_1 \hat{c}^\dagger (d-1/d) \cosh r \sinh r \right] \nonumber \\
& & - \mathcal{G} \cosh r \, \sinh r \left\{ \hat{\beta}_1 \hat{c}^\dagger \left[ e^{+2i\omega_2 t} - e^{-2i\omega_1 t} + \frac{1}{d} e^{+i(\omega_1 + \omega_2)t} - d e^{-i(\omega_1 + \omega_2)t} \right] \right. \nonumber \\
& & \ \ \ \ \ \ \ \ \ \ \ \ \ \ \ \ \ \ \ \ \ \ \left. + \hat{\beta}_2 \hat{c}^\dagger \left[ e^{+2i\omega_1 t} - e^{-2i\omega_2 t} + d e^{+i(\omega_1 + \omega_2)t} - \frac{1}{d} e^{-i(\omega_1 + \omega_2)t} \right] \right\} \nonumber \\
& & + \mathcal{G} \hat{\beta}^\dagger_1 \hat{c}^\dagger \left[ \cosh^2 r \left( e^{2i\omega_1 t} + d e^{i(\omega_1 + \omega_2)t} \right) - \sinh^2 r \left( \frac{1}{d} e^{-i(\omega_1 + \omega_2 )t} + e^{-2i\omega_2 t} \right) \right] \nonumber \\
& & + \mathcal{G} \hat{\beta}^\dagger_2 \hat{c}^\dagger \left[ \cosh^2 r \left( e^{2i\omega_2 t} + \frac{1}{d} e^{i(\omega_1 + \omega_2)t} \right) - \sinh^2 r \left( d e^{-i(\omega_1 + \omega_2 )t} + e^{-2i\omega_1 t} \right) \right] + {\rm h.c.} , \label{eq:CRfourBog}
\end{eqnarray}
\end{widetext}
where the asymmetry in the single-photon optomechanical coupling rates is parametrized by 
\begin{equation}
d \equiv \frac{g_a}{g_b}. 
\end{equation}

\subsubsection{Two-tone Driving}
With two driving tones, as per Eq.~(\ref{eq:twotone}), the time-dependent contribution to the Hamiltonian (\ref{eq:Htot}) is 
\begin{eqnarray}
\hat{\mathcal{H}}_{\rm CR} & = & G_+ \left( \hat{a} + \hat{b} \right) e^{-2i\omega_m t} \hat{c}^\dagger \nonumber \\ 
& & + G_- \left( \hat{a}^\dagger + \hat{b}^\dagger \right) e^{+2i\omega_m t} \hat{c}^\dagger + {\rm h.c.},  \label{eq:TwoCR}
\end{eqnarray}
only containing fast-rotating terms oscillating at $| 2\omega_m |$. 

\subsection{Drift matrix}
\label{sec:MatricesCR}
A time-dependent Hamiltonian (\ref{eq:Htot}) leads to a time-dependent drift matrix in the corresponding Heisenberg-Langevin equations, see Eq.~(\ref{eq:HLeqns}). The drift matrix takes the form given in Eq.~(\ref{eq:driftt}), and we specify the coefficient matrices here. In writing out these matrices it is useful to parameterize the coupling imperfection by $\varepsilon_{\pm}$, where
\begin{equation}
\frac{ \bar{c}_{1\pm} }{ \bar{c}_{2\pm} } \varepsilon_{\pm} \equiv \frac{g_b}{g_a} .
\end{equation}
Having both sets of drives matched according to Eq.~(\ref{eq:matchingcondition}), and therefore no imperfection in the effective coupling, corresponds to $\varepsilon_{\pm} = 1$. The drift coefficient matrices, for the general case of four-tone driving, are
\begin{subequations}
\begin{eqnarray}
\mathbf{A}_{1+} & = & \frac{1}{2} \left[ \begin{array}{c|c|c} \mathbf{0}_{22} & \mathbf{0}_{22} & d \tilde{\mathbf{M}}_+ \\ \hline \mathbf{0}_{22} & \mathbf{0}_{22} & \mathbf{N}_+/d \\ \hline d \tilde{\mathbf{M}}_- & \mathbf{N}_-/d & \mathbf{0}_{22} \end{array} \right] , \\
\mathbf{A}_{2+} & = & \frac{1}{2} \left[ \begin{array}{c|c|c} \mathbf{0}_{22} & \mathbf{0}_{22} & \mathbf{0}_{22} \\ \hline \mathbf{0}_{22} & \mathbf{0}_{22} & \tilde{\mathbf{Q}}_+ \\ \hline \mathbf{0}_{22} & \tilde{\mathbf{Q}}_- & \mathbf{0}_{22} \end{array} \right] , \\
\mathbf{A}_{3+} & = & \frac{1}{2} \left[ \begin{array}{c|c|c} \mathbf{0}_{22} & \mathbf{0}_{22} & d \tilde{\mathbf{Q}}_+ \\ \hline \mathbf{0}_{22} & \mathbf{0}_{22} & \mathbf{Q}_+/d \\ \hline d \tilde{\mathbf{Q}}_- & \mathbf{Q}_-/d & \mathbf{0}_{22} \end{array} \right] , \\
\mathbf{A}_{4+} & = & \frac{1}{2} \left[ \begin{array}{c|c|c} \mathbf{0}_{22} & \mathbf{0}_{22} & \tilde{\mathbf{Q}}_+ \\ \hline \mathbf{0}_{22} & \mathbf{0}_{22} & \mathbf{0}_{22} \\ \hline \tilde{\mathbf{Q}}_- &  \mathbf{0}_{22} & \mathbf{0}_{22} \end{array} \right] ,
\end{eqnarray}
\end{subequations}
where $\mathbf{A}_{k-} = \mathbf{A}^*_{k+}$, $\mathbf{0}_{22}$ is the $2 \times 2$ zero matrix and we have introduced the notation
\begin{subequations}
\begin{eqnarray}
\tilde{\mathbf{M}}_{\pm} & = & \left[ \begin{array}{cc} i(\mp \tilde{G}_- \tilde{\varepsilon}_+ - \tilde{G}_+ \tilde{\varepsilon}_- ) & \tilde{G}_- \tilde{\varepsilon}_+ - \tilde{G}_+ \tilde{\varepsilon}_- \\ - \tilde{G}_- \tilde{\varepsilon}_+ - \tilde{G}_+ \tilde{\varepsilon}_- & i ( \mp \tilde{G}_- \tilde{\varepsilon}_+ + \tilde{G}_+ \tilde{\varepsilon}_- ) \end{array} \right] , \nonumber \\
& & \\
\mathbf{N}_{\pm} & = & \left[ \begin{array}{cc} i(\pm G_- \tilde{\varepsilon}_+ + G_+ \tilde{\varepsilon}_- ) & G_- \tilde{\varepsilon}_+ - G_+ \tilde{\varepsilon}_- \\
-G_- \tilde{\varepsilon}_+ - G_+ \tilde{\varepsilon}_- & i (\pm G_- \tilde{\varepsilon}_+  - G_+ \tilde{\varepsilon}_- ) \end{array} \right] , \nonumber \\
& & 
\end{eqnarray}
\begin{eqnarray}
\tilde{\mathbf{Q}}_{\pm} & = & \left[ \begin{array}{cc} i(-\tilde{G}_- \tilde{\varepsilon}_+ \mp \tilde{G}_+ \tilde{\varepsilon}_- ) & -\tilde{G}_- \tilde{\varepsilon}_+ + \tilde{G}_+ \tilde{\varepsilon}_- \\ - \tilde{G}_- \tilde{\varepsilon}_+ - \tilde{G}_+ \tilde{\varepsilon}_- & i ( \tilde{G}_- \tilde{\varepsilon}_+ \mp \tilde{G}_+ \tilde{\varepsilon}_- ) \end{array} \right] , \nonumber \\
& & 
\end{eqnarray}
\end{subequations}
with $\mathbf{Q}_{\pm} $ given by $\tilde{\mathbf{Q}}_{\pm} $ with the replacement $\tilde{G}_{\pm} \rightarrow G_{\pm}$. The scalar tilde quantities are defined as
\begin{subequations}
\begin{eqnarray}
\tilde{G}_{\pm} & = & G_{\pm} \varepsilon_{\pm} , \\
\tilde{\varepsilon}_{\pm} & = & \frac{ 2(1 + \varepsilon_{\pm}) }{ (1 + \varepsilon_+ ) (1 + \varepsilon_-) } .
\end{eqnarray}
\end{subequations}

\subsection{Time-dependent drift matrix: solution}
\label{sec:SolnCR}
Given the form of the drift matrix (\ref{eq:driftt}), we expect the covariance matrix $\mathbf{V}$, given by the solution of (\ref{eq:LyapunovDE}), to be oscillatory in the long-time limit. We approximate the solution via the covariance matrix ansatz \cite{mari:nonRWA},
\begin{equation}
\mathbf{V}(t) = \mathbf{V}_0 + \sum^N_{k=1} \left( \mathbf{V}_{k+} e^{+2i\delta_k t} + \mathbf{V}_{k-} e^{-2i\delta_k t} \right) . \label{eq:CMAnsatz}
\end{equation}
In general, the solution will contain harmonics of the bare frequencies that appear in Eq.~(\ref{eq:driftt}), as well as their sum and difference frequencies. However, the solution that we really seek is the DC component of the covariance matrix, $\mathbf{V}_0$.

The equations of motion (\ref{eq:HLeqns}), with the drift matrix (\ref{eq:driftt}), may be written in the frequency domain as
\begin{eqnarray}
& & \sum^N_{k=1} \left( \mathbf{A}_{k+} \vec{X}[\omega - 2 \delta_k] + \mathbf{A}_{k-} \vec{X}[\omega + 2 \delta_k] \right) \nonumber \\ 
& & + (\mathbf{A}_0 - i\omega \mathbf{I}_6) \vec{X}[\omega ] = - \mathbf{B} \cdot \vec{X}_{\rm in} [ \omega ] \equiv \vec{N} [\omega ] , \nonumber \\
& & 
\end{eqnarray}
where we have defined the Fourier transform as $\mathcal{F} [\omega ] = \int^{+\infty}_{-\infty} f(t) \, e^{-i\omega t} \, dt$. Now we form the frequency-dependent state and noise vectors. These are $(2N+1)-$dimensional vectors where $N$ is the number of positive-frequency counter-rotating terms in Eq.~(\ref{eq:driftt}). The $n^{th}$ elements are $\vec{X} [ \omega - 2\delta_{| N+1-n |} \, {\rm sgn} \, (N+1-n) ] $ and $ \vec{N} [ \omega - 2 \delta_{| N+1-n |} \, {\rm sgn} \, (N+1-n)] $, respectively. Subsequently we can write the linear system 
\begin{equation}
\bar{\mathbf{A}}[ \omega ] \cdot \vec{\mathbf{X}} [ \omega ] = \vec{\mathbf{N}} [ \omega ] , \label{eq:finallinearsystem}
\end{equation}
where
\begin{widetext}
\begin{equation}
\bar{\mathbf{A}} [\omega ] = \left[ \begin{array}{ccccccc} \mathbf{A}_0 - i(\omega - 2 \delta_N)\mathbf{I}_6 & & & \mathbf{A}_{N-} & & & \\ & \ddots & & \vdots & & & \\ & & \ddots & \mathbf{A}_{1-} & & & \\ \mathbf{A}_{N+} & \ldots & \mathbf{A}_{1+} & \mathbf{A}_0 - i\omega \mathbf{I}_6 & \mathbf{A}_{1-} & \ldots & \mathbf{A}_{N-} \\ & & & \mathbf{A}_{1+} & \ddots & & \\ & & & \vdots & & \ddots & \\ & & & \mathbf{A}_{N+} & & & \mathbf{A}_0 - i(\omega + 2\delta_N)\mathbf{I}_6 \end{array} \right] .
\end{equation}
\end{widetext}

Introducing the noise correlation matrix $\Phi_{i,j} [ \omega , \omega' ] = \langle N_i [ \omega ] N^*_j [ \omega' ] + N^*_j [ \omega' ] N_i [ \omega ] \rangle/2$ leads to 
\begin{eqnarray}
\mathbf{\Phi} [ \omega , \omega' ] & = & \mathbf{D}_0 \, \delta [ \omega - \omega' ] + \sum^N_{k=1} \left( \mathbf{D}_{k+} \, \delta [ \omega - \omega' - 2\delta_k ] \right. \nonumber \\
& & \left. + \mathbf{D}_{k-} \, \delta [ \omega - \omega' + 2\delta_k ] \right) ,
\end{eqnarray}
where the matrices are defined by: $(\mathbf{D}_0)_{ii} = \mathbf{B} \mathbf{B}^T$, $(\mathbf{D}_{n+})_{N+1-n,N+1} = (\mathbf{D}_{n+})_{N+1,N+1+n} = \mathbf{B} \mathbf{B}^T$ for $n \in \left\{ 1, \ldots, N \right\}$, and $\mathbf{D}_{n-} = \mathbf{D}^T_{n+}$. The indices refer to $6\times 6$ blocks in the overall matrix. 

Solving the linear system (\ref{eq:finallinearsystem}) leads to 
\begin{equation}
\mathbf{V} [ \omega , \omega' ] = \bar{\mathbf{A}}^{-1} [ \omega ] \mathbf{\Phi} [ \omega,\omega' ] \left( \bar{\mathbf{A}}^{-1}[ \omega' ] \right)^\dagger .
\end{equation}
We really want the $6 \times 6$ central block which we denote $\tilde{\mathbf{V}}[ \omega, \omega' ]$. This is given by
\begin{eqnarray}
\tilde{\mathbf{V}} [ \omega , \omega' ] & = & \tilde{\mathbf{V}}_0 \delta [ \omega - \omega' ] \nonumber \\
& & + \sum^N_{k=1} \left( \tilde{\mathbf{V}}_{k+} [ \omega ] \delta [ \omega - \omega' - 2 \delta_k ] \right. \nonumber \\
& & \left. + \tilde{\mathbf{V}}_{k-} [ \omega ] \delta [ \omega - \omega' + 2 \delta_k ] \right) ,
\end{eqnarray}
where we have the coefficients
\begin{subequations}
\begin{eqnarray}
\tilde{\mathbf{V}}_0 [ \omega ] & = & \left[ \bar{\mathbf{A}}^{-1} [ \omega ] \mathbf{D}_0 \left( \bar{\mathbf{A}}^{-1} [\omega ] \right)^\dagger \right]_6 , \\
\tilde{\mathbf{V}}_{k\pm} [ \omega ] & = & \left[ \bar{\mathbf{A}}^{-1}[\omega ] \mathbf{D}_{k\pm} \left( \bar{\mathbf{A}}^{-1} [\omega \mp 2 \delta_k ] \right)^\dagger \right]_6 . \nonumber \\
& & 
\end{eqnarray}
\end{subequations}
The coefficients in Eq.~(\ref{eq:CMAnsatz}) follow from 
\begin{subequations}
\begin{eqnarray}
\mathbf{V}_0 & = & \frac{1}{2\pi} \int^{+\infty}_{-\infty} \tilde{\mathbf{V}}_0 [\omega ] d\omega, \\
\mathbf{V}_{k\pm} & = & \frac{1}{2\pi} \int^{+\infty}_{-\infty} \tilde{\mathbf{V}}_{k\pm} [\omega ] d\omega .
\end{eqnarray}
\end{subequations}

\end{document}